\documentclass[pdflatex,sn-basic, iicol]{sn-jnl}

\usepackage{graphicx}%
\usepackage{multirow}%
\usepackage{amsmath,amssymb,amsfonts}%
\usepackage{amsthm}%
\usepackage{mathrsfs}%
\usepackage[title]{appendix}%
\usepackage{xcolor}%
\usepackage{textcomp}%
\usepackage{manyfoot}%
\usepackage{booktabs}%
\usepackage{algorithm}%
\usepackage{algorithmicx}%
\usepackage{algpseudocode}%
\usepackage{listings}%
\usepackage{subfigure}
\usepackage{placeins}

\theoremstyle{thmstyleone}%
\theoremstyle{thmstyletwo}%

\theoremstyle{thmstylethree}%

\begin{document}

\title[Efficient Uncertainty Propagation in Bayesian Two-Step Procedures]{Efficient Uncertainty Propagation in Bayesian Two-Step Procedures}

\author*[1]{\fnm{Svenja} \sur{Jedhoff}}\email{jedhoff@statistik.tu-dortmund.de}

\author[2]{\fnm{Hadi} \sur{Kutabi}}\email{hadi.kutabi@kit.edu}

\author[2]{\fnm{Anne} \sur{Meyer}}\email{anne.meyer@kit.edu}

\author[1]{\fnm{Paul-Christian} \sur{Bürkner}}\email{buerkner@statistik.tu-dortmund.de}

\affil*[1]{\orgdiv{Department of Statistics}, \orgname{TU Dortmund University}, \orgaddress{\city{Dortmund}, \country{Germany}}}

\affil[2]{\orgdiv{Research Center for Information Technology}, \orgname{Karlsruhe Institute of Technology}, \orgaddress{\city{Karlsruhe}, \postcode{76131}, \country{Germany}}}

\abstract{Bayesian inference provides a principled framework for probabilistic reasoning. If inference is performed in two steps, uncertainty propagation plays a crucial role in accounting for all sources of uncertainty and variability. This becomes particularly important when both aleatoric uncertainty, caused by data variability, and epistemic uncertainty, arising from incomplete knowledge or missing data, are present. Examples include surrogate models and missing data problems. In surrogate modeling, the surrogate is used as a simplified approximation of a resource-heavy and costly simulation. The uncertainty from the surrogate-fitting process can be propagated using a two-step procedure. For modeling with missing data, methods like Multivariate Imputation by Chained Equations (MICE) generate multiple datasets to account for imputation uncertainty. These approaches, however, are computationally expensive, as multiple models must be fitted separately to surrogate parameters or imputed datasets.
To address these challenges, we propose an efficient two-step approach that reduces computational overhead while maintaining accuracy. By selecting a representative subset of draws or imputations, we construct a mixture distribution to approximate the desired posteriors using Pareto smoothed importance sampling. For more complex scenarios, this is further refined with importance weighted moment matching and an iterative procedure that broadens the mixture distribution to better capture diverse posterior distributions.}


\maketitle

\section{Introduction}

Many modeling approaches in science and engineering adopt a two-step framework, wherein an initial modeling or preprocessing phase is followed by a subsequent stage focused on inference or prediction \citep{liu_current_2021, lim_enhancing_2024, kim_two-step_2020, shin_time_2023, cacao_machine_2025}. While effective, such two-step procedures introduce additional complexity --- particularly in the treatment of uncertainty. A key challenge in two-stage inference lies in accurately propagating uncertainty across steps to ensure the reliability and trustworthiness of the final outcomes \citep{smith_uncertainty_2013, peherstorfer_survey_2018}. Addressing this challenge often entails computationally intensive and time-consuming procedures.
In this paper, we propose a general methodology for efficient and robust uncertainty propagation within Bayesian two-step modeling frameworks. Here, robustness refers to the ability of the method to diagnose and address unreliable importance-sampling approximations, especially when the posterior distributions arising in the two steps differ substantially. The proposed method uses the PSIS diagnostic $\hat{k}$ to detect poor overlap between proposal and target distributions and applies IWMM as an adaptive correction when necessary. We demonstrate the effectiveness of our approach in two widely encountered statistical scenarios that demand careful uncertainty management, as described in the following sections.

\begin{figure*}[ht]
    \centering
    \includegraphics[width=\linewidth]{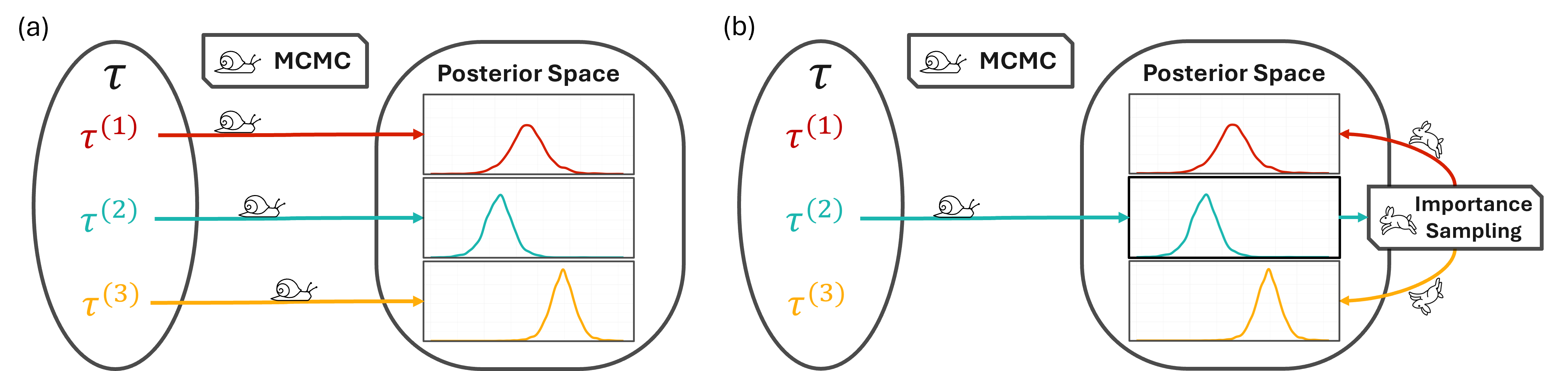}
    \caption{Overview of problem setting: The goal is to propagate the uncertainty induced by multiple draws of $\tau$ into the posterior space.
    \textbf{(a)} Standard approach: Run costly simulation, i.e. MCMC, for each realization of $\tau$.
    \textbf{(b)} Proposed method in this paper: Only a few costly simulations, i.e. MCMC, are required; the remaining posterior distributions are approximated using importance sampling.}
    \label{fig:introduction}
\end{figure*}

A prominent example of a two-step modeling approach is the use of surrogate models in scenarios where simulations of complex phenomena are required --- such as in the natural sciences \citep{koziel_surrogate-based_2013, forrester_engineering_2008}. These simulations often involve solving differential equations that lack closed-form solutions and must be addressed through computationally expensive numerical methods \citep{quarteroni_numerical_2007}. 
To alleviate these computational challenges, surrogate models \citep{zhu_bayesian_2018, lavin_simulation_2022} are employed as simplified approximations of the true system, enabling significantly faster evaluations while preserving essential behaviors of the original model.
Surrogate modeling has found broad application across domains such as systems biology \citep{renardy_parameter_2018, alden_using_2020}, water resource management \citep{razavi_review_2012, garzon_machine_2022}, and climate modeling \citep{kuehnert_surrogate_2022}.
In a typical two-step surrogate modeling approach \citep{reiser_uncertainty_2025}, the complex simulator is first approximated using a simpler surrogate trained on a limited set of outputs from the true model. This introduces uncertainty stemming from both limited training data and potential model mismatch. In the second step, inference is performed using the trained surrogate to estimate parameters based on observed data. To ensure reliable parameter estimates, it is critical to account for and appropriately propagate the uncertainties introduced during surrogate construction into the inference stage \citep{psaros_uncertainty_2023}.

Another context in which a two-step procedure is essential arises when working with datasets containing missing values. Rather than discarding incomplete records --- which may lead to biased results or information loss --- imputation techniques are employed to estimate the missing entries \citep{buuren_flexible_2018}. Among these, Multivariate Imputation by Chained Equations (MICE) \citep{van_buuren_flexible_1999} stands out as a widely adopted method \citep{rubin_multiple_2004, white_multiple_2011, shah_comparison_2014}.
MICE addresses uncertainty from missingness by generating multiple complete datasets, each containing plausible imputations for the missing values \citep{van_buuren_flexible_1999}. In the second step, a statistical model is independently fitted to each imputed dataset, and the resulting posterior distributions are combined. While this approach effectively incorporates imputation uncertainty, it can become computationally demanding, especially when complex models are used. Nevertheless, it yields more robust and comprehensive posterior estimates by properly accounting for the uncertainty introduced during imputation.

In both examples --- surrogate modeling and missing data handling --- the key challenge is propagating the uncertainty from the first to the second step. A major drawback of existing methods is their high computational cost, as they require separate model fits for each realization of the uncertain quantity (e.g. surrogate parameters or imputed datasets). This issue is particularly acute in the Bayesian setting, where posterior distributions are often approximated using costly sampling algorithms \citep{gelman_bayesian_2013, robert_monte_2004}, most notably Markov Chain Monte Carlo (MCMC) \citep{brooks_handbook_2011}. 

A closely related line of work arises in cut models (or cutting feedback), where inference is modularized to prevent information from a downstream (potentially misspecified) module from feeding back into an upstream module \citep{bayarri_modularization_2009, lunn_bugs_2009, plummer_cuts_2015}. Therefore, cut models can be viewed as Bayesian two-step procedures and, in this sense, constitute a special case of our framework.
In practice, most cut methods fall into two broad classes. 
First, two-step Monte Carlo approaches, similar in spirit to multiple imputation, propagate uncertainty by repeatedly drawing a realization $\tau^{(i)}$ of the uncertain quantity obtained in the first step and then refitting the second-step Bayesian model conditional on that draw \citep{plummer_cuts_2015}.
Second, tailored algorithms approximate the cut distribution more directly, for example via tempered or modified transition kernels (motivated in part by the failure of naive cut samplers to target a well-defined stationary distribution) \citep{plummer_cuts_2015}, stochastic-approximation methods such as SACut \citep{liu_stochastic_2021}, or semi-modular approaches that downweight feedback rather than removing it entirely \citep{carmona_semi-modular_2020}. 
Beyond MCMC-based implementations, alternative strategies include posterior-bootstrap procedures for robust modular inference \citep{pompe_asymptotics_2021}, generalized-Bayes formulations of cutting feedback \citep{frazier_cutting_2025}, Gaussian-process emulation of conditional posteriors to reduce the number of costly refits \citep{hutchings_enhancing_2025}, sequential Monte Carlo for cut posterior computation \citep{mathews_sequential_2025}, and variational approaches that cast cutting feedback as constrained optimization \citep{yu_variational_2023, song_neural_2025}. 
Unbiased MCMC estimators based on couplings provide another general tool for estimating expectations under such targets \citep{jacob_unbiased_2020}. 
While these methods address important conceptual and computational challenges, many of them still require repeated downstream posterior computations or substantial additional approximation effort. This motivates the development of efficient uncertainty propagation strategies for Bayesian two-step workflows. In Appendix~\ref{sec:ap:nevicut}, we therefore include an illustrative comparison with the neural variational inference approach to cutting feedback proposed by \citet{song_neural_2025}. In the missing-data simulation setting considered there, our PSIS+IWMM-based method yields posterior summaries that are closer to the MCMC baseline than those obtained with the so-called NeVI-Cut approach of \citet{song_neural_2025}.

For example, if 100 realizations $\tau^{(1)},\dots,\tau^{(100)}$ are generated in the first step, this would typically necessitate estimating 100 separate models --- each costly in terms of time and computational resources (see Figure \ref{fig:introduction}(a)). To overcome this limitation, we propose an efficient approach that significantly reduces the computational overhead while preserving the trustworthiness of uncertainty propagation. 
Our approach leverages importance sampling techniques, allowing us to approximate the desired posterior distributions without running MCMC or comparably costly algorithms for each realization $\tau^{(i)}$. Specifically, we employ Pareto Smoothed Importance Sampling (PSIS) \citep{vehtari_pareto_2024} and Importance Weighted Moment Matching (IWMM) \citep{paananen_implicitly_2021}. 
With this strategy, MCMC is required only for a small subset of realizations of $\tau$, reducing computational demands. For the remaining realizations, we approximate the corresponding posterior distributions directly within the posterior space using advanced importance sampling (see Figure \ref{fig:introduction}(b)).
This approach offers a significant gain in efficiency while preserving the accuracy of the propagated uncertainty, making it a practical scalable solution for complex two-step modeling workflows.

\section{Method}

In this section, we present our proposed methodology for addressing uncertainty propagation within Bayesian two-step modeling frameworks. We begin by formally defining the problem setting and introducing the key components required for our approach. Section \ref{sec:iterative_method} then describes the core of our method --- an iterative procedure designed for efficient and reliable uncertainty propagation across the two stages. Finally, we discuss two specific application scenarios that require additional notation and domain-specific considerations.

\subsection{Problem Definition}
The primary objective is to obtain information about the distribution of a variable of interest $ \theta \sim p(\theta \mid C) $, where $C$ denotes additional context, which depends on the specific problem domain. The marginalized posterior $p(\theta \mid C)$ can be obtained by integrating out the underlying variable $ \tau $ via 
\begin{align}  
    p(\theta \mid C) = \int p(\theta \mid \tau, C) p(\tau \mid C) \, d\tau .  
    \label{eq:problem}  
\end{align}  
Here, the probability distribution $ p(\tau \mid C) $ is either unknown or intractable, making the direct computation of the above integral infeasible.  
If we have independent values $ \{\tau^{(i)} \}_{i=1}^m \sim p(\tau \mid C) $, we can approximate the above integral with a Monte Carlo estimator \citep{hammersley_monte_1964} as
\begin{align}
    \int p(\theta \mid \tau, C) p(\tau \mid C) \, d\tau \approx \frac{1}{m} \sum_{i=1}^m p\left(\theta \mid \tau^{(i)}, C\right).  
    \label{eq:mc_estimate}
\end{align}
By the law of large numbers, we obtain convergence for $ m \to \infty $ \citep{robert_monte_2004}. However, even for finite but sufficiently large $ m $, we can achieve a reasonable approximation of the marginalized posterior $ p(\theta \mid C) $ \citep{bishop_pattern_2006}. In the following, we simplify notation by omitting the explicit dependence of $p(\theta \mid \tau, C)$ on the context variable $C$. As a result, $p(\theta \mid \tau)$ should be understood as implicitly conditioned on $C$, even though it is not explicitly shown. For a clearer understanding of the role and interpretation of the context $C$, refer to the two application domains discussed in Sections \ref{sec:special_case:MI} and \ref{sec:special_case:surrogate}.

In many practical scenarios, the number of first-step realizations $m$ available for uncertainty propagation is limited by the cost of generating or representing uncertainty in the first step. In the multiple-imputation setting, the number of imputed datasets is finite and is typically chosen as a compromise between accurately representing missing-data uncertainty and computational feasibility.
Additionally, evaluating $ p(\theta\mid\tau^{(i)}) $ for each $ i=1, \dots, m $ can be computationally expensive \citep{neal_probabilistic_1993}. In the setting of Equation~(\ref{eq:mc_estimate}), the posterior distribution of $\theta$ given a $\tau^{(i)}$, $p(\theta\mid\tau^{(i)}) $, cannot be accessed directly and must instead be approximated through expensive sampling algorithms \citep{gelman_bayesian_2013}, for example, via MCMC \citep{robert_monte_2004}. For each $ \tau^{(i)} $, we then obtain $ S $ samples $\{\theta^{(s)}_i\}_{s=1}^S \sim p(\theta\mid\tau^{(i)}) $ from the corresponding posterior distribution.
By treating all $ m \cdot S $ draws as one large sample, we can use them to approximate the marginalized posterior distribution $ {p(\theta \mid C)} $ \citep{lindsay_mixture_1995}. 

This approach can be computationally expensive, depending on the complexity of the distributions involved. A straightforward alternative to running MCMC $ m $ times is to select a single representative value $ \tau^* $ from the set of available values $ \{\tau^{(i)}\}_{i=1}^m $ and perform MCMC only for this value:
\begin{align}  
    p(\theta \mid C) = \int p(\theta \mid \tau) p(\tau \mid C) \, d\tau \approx p(\theta \mid \tau^*) \, .  
\end{align} 
To ensure $ \tau^* $ is representative of the entire sample, it can be chosen as a summary statistic, such as the mean or median of the $ m $ samples, depending on the characteristics of $ \tau $. While this approach is computationally efficient --- requiring only a single MCMC run --- it fails to propagate the uncertainty in $ p(\tau \mid C) $, almost surely leading to overconfidence in the resulting marginalized posterior distribution $ p(\theta \mid C) $. 

A more accurate approximation of $ p(\theta \mid C) $ requires using all available samples via the Monte Carlo estimate in Equation \eqref{eq:mc_estimate}. To make this approach computationally feasible, the number of required MCMC runs needs to be reduced. To this end, we propose a combination of importance sampling techniques --- specifically, Pareto Smoothed Importance Sampling (PSIS) \citep{vehtari_pareto_2024} and importance weighted moment matching (IWMM) \citep{paananen_implicitly_2021}, which are described in detail in Section \ref{sec:method_components}. These methods allow us to approximate the distributions $ {p(\theta \mid \tau^{(i)})} $ for $ \{\tau^{(i)}\}_{i=1}^m $ within the posterior space as shown in Figure \ref{fig:introduction}(b). This facilitates a more efficient calculation of the Monte Carlo estimate in Equation~(\ref{eq:mc_estimate}) than running costly sampling algorithms for each value of $ \{\tau^{(i)}\}_{i=1}^m $ separately (see Figure \ref{fig:introduction}(a) for comparison).

In the following, we introduce the key components required to enhance the computation of the Monte Carlo estimator. These components are then integrated in Section~\ref{sec:iterative_method} to construct an efficient iterative algorithm.

\subsection{Method Components \label{sec:method_components}}

Our objective is to approximate the distribution $p(\theta \mid \tau^{(i)})$ without relying on computationally expensive sampling algorithms. In the following, for simplicity, we refer only to MCMC to describe this class of expensive but highly accurate algorithms. Instead of running MCMC to access the distribution, we utilize a proposal distribution $p(\theta \mid \tau^*)$ and apply PSIS. When PSIS alone is insufficient to accurately approximate $p(\theta \mid \tau^{(i)})$, we enhance the approximation using IWMM.

\textbf{Importance Sampling}\quad 
The key idea behind importance sampling is to approximate properties of the target distribution $p(\theta\mid\tau^{(i)})$ without drawing samples directly from it. This allows, among other things, to approximate expectations $\mathbb{E}_{p(\theta|\tau^{(i)})}(h(\theta)\mid\tau^{(i)})$ over the target for any function $h$ \citep{hammersley_monte_1964}.  

Rather than sampling directly from the target, a proposal distribution, $p(\theta\mid\tau^*)$ in our framework, is used as an initial approximation. Here, $\tau^* \in \{\tau^{(i)} \}_{i=1}^m$ is a selected value, for which we access the corresponding posterior $p(\theta \mid \tau^*)$ using MCMC. In the following, we use self-normalizing importance sampling, since both target and proposal densities are only known up to a constant. The expectation of interest can then be expressed as follows \citep{gelman_bayesian_2013}:  
\begin{align}
    \mathbb{E}_{p(\theta \mid \tau^{(i)})}(h(\theta)\mid\tau^{(i)}) = \frac{\int h(\theta) p(\theta\mid\tau^{(i)}) d\theta }{\int p(\theta\mid\tau^{(i)}) d\theta } \\
=   \frac{\int h(\theta) \frac{p(\theta\mid\tau^{(i)})}{p(\theta\mid\tau^*)}p(\theta\mid\tau^{*}) d\theta }{\int \frac{p(\theta\mid\tau^{(i)})}{p(\theta\mid\tau^*)}p(\theta\mid\tau^{*}) d\theta } \, .
\end{align}
That is, we express expectations over the target equivalently as expectations over the proposal. Then, given samples $\{\theta^{(s)}_*\}_{s=1}^S \sim p(\theta\mid\tau^*)$ drawn from the proposal, the desired target expectation can be estimated as
\begin{align}
\mathbb{E}_{p(\theta|\tau^{(i)})}(h(\theta)\mid\tau^{(i)}) &\approx  \frac{\sum_{s=1}^S h(\theta^{(s)}_*) r_i(\theta^{(s)}_*) }{ \sum_{s=1}^S r_i(\theta^{(s)}_*)}  \\
&= \sum_{s=1}^S h(\theta^{(s)}_*) w_i(\theta^{(s)}_*), \\
\text{  with  } \; r_i(\theta^{(s)}_*) &= \frac{p(\theta^{(s)}_*\mid\tau^{(i)})}{p(\theta^{(s)}_*\mid\tau^*)} \label{eq:importance_ratios} \; \\
\text{  and  } \; w_i(\theta^{(s)}_*) &= \frac{r_i(\theta^{(s)}_*)}{\sum_{s=1}^S r_i(\theta^{(s)}_*)} \,, 
\end{align}
$s=1, \dots, S$. Here, $r_i(\theta^{(s)}_*)$, for $s=1, \dots, S$, are referred to as importance ratios and $w_i(\theta^{(s)}_*)$ are the corresponding importance weights. 
Using the same proposal distribution $p(\theta \mid \tau^*)$, we can compute importance weights for each of the posterior distributions $p(\theta \mid \tau^{(i)})$ corresponding to the values $\tau^{(i)} \in \{\tau^{(i)}\}_{i=1}^m$. For importance sampling to be effective and yield reliable approximations, the target and proposal distributions must be sufficiently similar and share the same support.
Since both $\tau^{(i)}$ and $\tau^*$ are drawn from the same distribution $p(\tau \mid C)$, the corresponding posteriors $p(\theta \mid \tau^{(i)})$ and $p(\theta \mid \tau^*)$ typically share similar structure and support, which enables us to apply importance sampling effectively.

Instead of computing expectations directly, the resulting importance weights can also be used to generate approximate samples from the target distributions via weighted resampling of the original draws. We apply this method, known as importance resampling \citep{gelman_bayesian_2013}, to obtain random draws from each of the posterior distributions $p(\theta \mid \tau^{(i)})$ for $i = 1, \dots, m$.

\textbf{Pareto Smoothed Importance Sampling}\quad 
Importance sampling, and consequently importance resampling, can be unreliable if the proposal distribution is not well-suited for the target distribution \citep{bugallo_adaptive_2017}. This issue can be diagnosed by examining the distribution of the importance weights $\{w_i(\theta^{(s)}_*)\}_{s=1}^S$. If the distribution appears to have excessively heavy tails, importance sampling becomes unstable, often with very large or even infinite variance, leading to unreliable estimates \citep{ionides_truncated_2008, vehtari_pareto_2024}. To address this problem, various adaptations of importance sampling have been developed \citep{ionides_truncated_2008, neal_annealed_2001}. In the following, we apply PSIS \citep{vehtari_pareto_2024}, since it is a reliable and trustworthy variant of importance sampling that outperforms other methods with regard to stabilizing importance sampling estimates.

In this approach, the importance weights are first sorted in ascending order. A Pareto distribution is then fitted to the largest ${M = \lfloor \min(0.2S, 3\sqrt{S})\rfloor}$ of the $S$ importance weights. These $M$ largest weights are subsequently replaced with the corresponding quantiles from the fitted generalized Pareto distribution. Beyond improving the stability of the importance weights, PSIS also provides a diagnostic metric $\hat{k}$, to assess the reliability of importance sampling. Performance is evaluated using a threshold, computed as $k_\text{threshold} = \min(1 - 1/\log_{10}(S), 0.7)$. For large sample sizes ($S > 2000$), $k_\text{threshold}  = 0.7$ is sufficient, which is adopted here for simplicity \citep{vehtari_pareto_2024}.  

If $\hat{k}_i < 0.7$, the Pareto-smoothed importance weights $w_i$ are considered reliable and can be used to resample from $\{\theta^{(s)}_*\}_{s=1}^S$ to obtain ${\{\theta^{(s)}_i\}_{s=1}^S \sim p(\theta\mid\tau^{(i)})}$. 
However, when $\hat{k}_i \geq 0.7$, the proposal distribution is deemed insufficiently close to the target distribution for reliable importance sampling. In this case, the importance weights exhibit heavy-tailed behavior, and the Pareto smoothing can no longer adequately control the estimator's error characteristics. Specifically, the bias introduced by smoothing the extreme weights begins to dominate over the estimator's variance, undermining both the accuracy and interpretability of the resulting estimates \citep{vehtari_pareto_2024}. As a result, the PSIS-based approximation may be unstable or biased, and a more complex and advanced method might be necessary to successfully approximate the target distribution. To address this, we employ importance-weighted moment matching (IWMM), an adaptive importance sampling method described in detail below.

\textbf{Importance Weighted Moment Matching}\quad
The method introduced by \cite{paananen_implicitly_2021} can be employed when PSIS fails, as indicated by a $\hat{k} > k_\text{threshold} $. Importance weighted moment matching (IWMM) is an adaptive importance sampling method, where the proposal distribution is iteratively updated to better approximate the target distribution. In general, these methods consist of three steps: (i) generating draws from the proposal distribution, (ii) computing the importance weights for these draws, and (iii) adapting the proposal distribution based on the computed weights \citep{bugallo_adaptive_2017}.  

In the specific case of IWMM, step (i) is omitted, so that the same samples are transformed directly and no resampling is necessary. With these samples, the importance weights are calculated in step (ii). In step (iii), the samples are transformed using an affine transformation. While these transformations can vary in complexity, for simplicity, we focus on basic affine transformations, since both the transformation and its Jacobian are computationally cheap \citep{paananen_implicitly_2021}. 
To approximate the expectation $\mathbb{E}_{p(\theta \mid \tau^{(i)})}(h(\theta)\mid\tau^{(i)})$ with a set of draws $\{\theta^{(s)}_*\}_{s=1}^S$ from the proposal distribution $p(\theta\mid\tau^*)$, we use a matrix $\mathbf{A}$ to represent a linear map and a vector $\mathbf{b}$ to represent a translation. A draw $\theta^{(s)}_*$ is then transformed as follows \citep{paananen_implicitly_2021}:  
\begin{align}
    T: \theta^{(s)}_* \mapsto \mathbf{A} \theta^{(s)}_*+\mathbf{b} =: \breve{\theta}^{(s)}_* \, .
\end{align}
Since the transformation is affine and the same for all draws, a new implicit density $p_T(\theta\mid\tau^*)$ is given by\begin{align}
    p_T(\breve{\theta}^{(s)}_*\mid\tau^*) = p(\theta^{(s)}_*\mid\tau^*) \; \cdot \mid\mathbf{J}_T\mid^{-1} \, ,
\end{align}
where $ \mid\mathbf{J}_T\mid^{-1}$ is the inverse Jacobian $ \mid\mathbf{J}_T\mid^{-1} = \left|\frac{\mathrm{d}T(\theta)}{\mathrm{d}\theta}\right|^{-1}$.
To be more specific, three transformations are used to reduce the mismatch between the samples $\{\theta^{(s)}_*\}_{s=1}^S$ and the target distribution. Transformation $T_1$ is used to match the mean of the samples to the importance weighted mean.  This mean-shift transformation is closely related to the recentered importance sampling approach of \cite{mcvinish_recentered_2013}, which also improves importance sampling approximations by translating an existing Monte Carlo sample toward the target distribution. IWMM extends this idea by embedding the mean shift in a broader moment-matching framework that can additionally match marginal variances (with transformation $T_2$) or the full covariance structure (with transformation $T_3$) through affine transformations \citep{paananen_implicitly_2021}. The transformation $T_1$ to match the mean is given by:
\begin{align}
    \breve{\theta}^{(s)}_* &= T_1(\theta^{(s)}_*) = \theta^{(s)}_*  - \bar{\theta}_* + \tilde{\theta}_* \\ &\text{ with } \bar{\theta}_* = \frac{1}{S} \sum_{s=1}^S \theta^{(s)}_* \\ &\text{ and } \tilde{\theta}_* =\frac{\sum_{s=1}^S w_i( \theta^{(s)}_*) \theta^{(s)}_*}{\sum_{s=1}^S w_i(\theta^{(s)}_*)} \, .
\end{align}
Transformation $T_2$ is used to match the marginal variance in addition to matching the mean: 
\begin{gather}
    \breve{\theta}^{(s)}_* = T_2(\theta^{(s)}_*) = \tilde{\mathbf{v}}^{1/2} \circ \mathbf{v}^{-1/2} \circ (\theta^{(s)}_* - \bar{\theta}_*) + \tilde{\theta}_* \\ 
    \text{ with } \mathbf{v} = \frac{1}{S} \sum_{s=1}^S (\theta^{(s)}_* - \bar{\theta}_*) \circ (\theta^{(s)}_* - \bar{\theta}_*)\\ \text{ and } \tilde{\mathbf{v}} = \frac{\sum_{s=1}^S w_i(\theta^{(s)}_*) (\theta^{(s)}_* - \bar{\theta}_*) \circ (\theta^{(s)}_* - \bar{\theta}_*) }{\sum_{s=1}^S w_i(\theta^{(s)}_*)} \, ,
\end{gather}
with $\circ$ indicating the pointwise product of two vectors. To also match the covariance and the mean, transformation $T_3$ can be applied:

\begin{gather}
    \breve{\theta}^{(s)}_* = T_3(\theta^{(s)}_*) = \tilde{\mathbf{L}} \mathbf{L}^{-1} (\theta^{(s)}_* - \bar{\theta}_*) + \tilde{\theta}_* \\
    \text{ with } \mathbf{LL}^T = \mathbf{\Sigma} = \frac{1}{S} \sum_{s=1}^S (\theta^{(s)}_* - \bar{\theta}_*)(\theta^{(s)}_* - \bar{\theta}_*)^T \, , \\ \tilde{\mathbf{L}}\tilde{\mathbf{L}}^T = \frac{\sum_{s=1}^S w_i(\theta^{(s)}_*) (\theta^{(s)}_* - \tilde{\theta}_*) (\theta^{(s)}_* - \tilde{\theta}_*)^T }{\sum_{s=1}^S w_i(\theta^{(s)}_*)} \, .
\end{gather}

The affine transformations are applied in order of increasing complexity, requiring a larger effective sample size to accurately compute the moments for more complex transformations. These transformations are performed iteratively, and for each transformation new weights are computed as well as the corresponding $\hat{k}$ diagnostic from PSIS, now called $\hat{k}_{\rm MM}$ in the context of IWMM.
A transformation is accepted if the new $\hat{k}_{\rm MM}$ is lower than the previous $\hat{k}_{\rm MM}$. If a transformation does not improve $\hat{k}_{\rm MM}$, the next transformation in the sequence is attempted. The algorithm terminates successfully when $\hat{k}_{\rm MM}$ falls below $k_\text{threshold}$. 
However, if the most complex transformation is not able to improve $\hat{k}_{\rm MM}$ anymore, but $\hat{k}_{\rm MM}$ still exceeds $k_\text{threshold} $, the algorithm terminates by returning the results of the latest successful transformation and a warning about sampling inaccuracy. In this case, we consider IWMM unsuccessful \citep{paananen_implicitly_2021}.

\begin{figure*}[th]
    \centering
    \includegraphics[width=\linewidth]{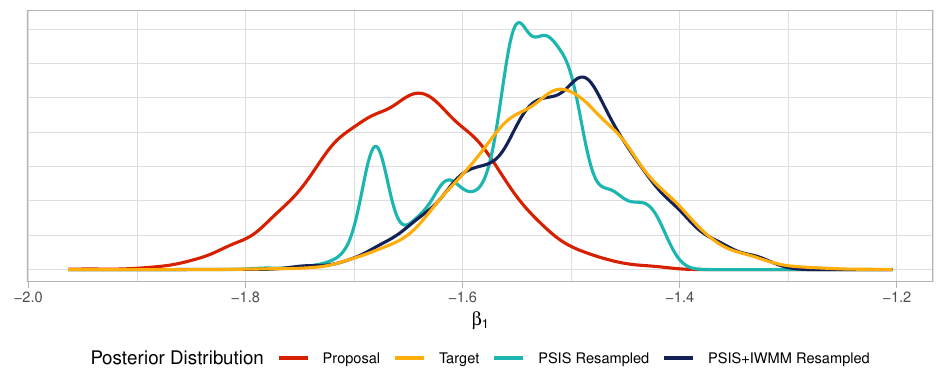}    
    \caption{Exemplary posterior distributions for one model parameter in the special case of the missing data problem (cf. Section \ref{sec:special_case:MI}) from the simulation study in Section \ref{sec:MI_sim_study} (dataset with $N=100$ observations, $p=10$ predictors and 15\% of rows containing missing values; imputed using MICE). The proposal and target distributions correspond to each one imputed dataset (each approximated with HMC). PSIS alone fails with a $\hat{k} = 0.93 > k_{\rm threshold}$, which implies an unsuccessful approximation of the target distribution as is clearly visible in the figure. In contrast, the combination of PSIS and IWMM appropriately approximates the target distribution, as indicated by $\hat{k}_{\rm MM} = 0.27 < k_{\rm threshold}$.}
    \label{fig:posteriors}
\end{figure*}

Although IWMM is computationally more expensive than PSIS, we employ it in cases where PSIS fails to provide reliable approximations of the target distribution. Figure~\ref{fig:posteriors} illustrates representative posterior distributions from two imputed datasets in the simulation study described in Section~\ref{sec:MI_sim_study}, within the special case of the missing data problem (see Section~\ref{sec:special_case:MI}). The proposal distribution (shown in red) was obtained via MCMC, and the true target distribution (yellow) --- which is typically unknown without running MCMC --- serves as the reference.
The PSIS-based approximation (light blue), constructed via importance resampling from the proposal, fails to accurately capture the shape and location of the target distribution. This poor approximation is flagged by a high Pareto shape parameter,  ${\hat{k} = 0.93 > k_{\mathrm{threshold}}}$, indicating that the results are unreliable. In contrast, additionally using IWMM results in a substantially improved approximation (dark blue), with resampled draws that closely match the target distribution. This is confirmed by a much lower shape parameter, ${\hat{k}_{\mathrm{MM}} = 0.27 < k_{\mathrm{threshold}}}$, indicating a successful and trustworthy approximation.

\textbf{Selection of Representative Values}\quad
To apply both PSIS and IWMM, we require a suitable proposal distribution along with corresponding samples. To obtain such a proposal, we run MCMC to reliably approximate one of the posterior distributions and generate representative draws from it. Specifically, we consider all values $\{\tau^{(i)}\}_{i \in I}$ for which the posterior $p(\theta \mid \tau^{(i)})$ has not yet been approximated. Initially, this set is defined as $I = \{1, \dots, m\}$.  The objective is to select one or more values from $\{\tau^{(i)}\}_{i\in I}$ such that their corresponding posterior distributions $p(\theta \mid \tau^{(i)})$ sufficiently represent the overall space of posteriors. By choosing such representative posteriors as proposals, we aim to maximize the effectiveness of both PSIS and IWMM.
Depending on the characteristics of $\tau$, different selection strategies can be applied. These include randomly selecting a subset of values, computing summary statistics (such as the mean or median), or using a similarity or distance-based criterion \citep{stolte_methods_2024, kaufman_finding_1990}. Since the choice of strategy is highly problem-specific, further details are provided in Section~\ref{sec:special_case:MI} for the case of working with missing data, and in Section~\ref{sec:special_case:surrogate} for surrogate modeling scenarios.
A more general, problem-independent strategy involves selecting the value in $\{\tau^{(i)}\}_{i \in I}$ associated with the largest $\hat{k}$ or $\hat{k}_{\rm MM}$ from the previous iteration of the algorithm described in Section~\ref{sec:iterative_method}. In each iteration, a new proposal distribution is used to apply PSIS and IWMM, targeting the posterior distributions corresponding to the remaining values $\tau^{(i)} \in \{\tau^{(i)}\}_{i \in I}$.

To ensure progress in the iterative algorithm described in Section~\ref{sec:iterative_method}, we restrict the selection of the representative to the existing set of unexplored values, i.e. $\tau^* \in \{\tau^{(i)}\}_{i \in I}$. This guarantees that the size of the index set $I$ is reduced in each iteration, thus avoiding repetition and eventual stagnation of the algorithm.
When choosing a selection strategy like a summary statistic, this might not be the case. Let $\bar{\tau}$ be the summary statistic, for example the mean, of the set $\{\tau^{(i)}\}_{i\in I}$. Then $\bar{\tau}$ is not necessarily part of the set: $\bar{\tau} \notin \{\tau^{(i)} \}_{i\in I} $. It is possible that when choosing $p(\theta \mid \bar{\tau})$ as a proposal for the posteriors $p(\theta\mid \tau^{(i)})$ for $i \in I$, both PSIS and IWMM fail to produce successful approximations for any of the posteriors. In the next iteration a new proposal is chosen to try PSIS and IWMM again. But since no posterior was approximated successfully, the index set $I$ is the same and therefore the value selected for the proposal is the same summary statistic $\bar{\tau}$. This leads to an infinite loop and must therefore be avoided.

A single MCMC run is sufficient to fit a proposal distribution $p(\theta \mid \tau^*)$, which can then be used to approximate the posteriors corresponding to all values $\{\tau^{(i)}\}_{i \in I}$ using either PSIS or IWMM. Alternatively, instead of selecting a single value $\tau^*$ to construct the proposal, multiple representative values can be chosen, each requiring a separate MCMC run to approximate its corresponding posterior distribution.
Although this approach incurs a higher computational cost, it offers notable advantages: combining the resulting posterior approximations yields a mixture distribution that is broader and more effectively captures the diversity across the posterior space. Such a mixture serves as an improved proposal distribution for PSIS, as it enhances coverage and mitigates the risk of poor overlap with the target distributions.
The detailed construction of this mixture distribution is presented in the following section.

\textbf{Mixture Proposal Distributions \label{sec:mixture}}\quad
To construct a mixture distribution for use as a proposal in PSIS, multiple representative values must be selected from the set $\{\tau^{(i)}\}_{i \in I}$. Let $\{\tau^{(j)}\}_{j \in I_{\text{mix}}}$ denote this selected subset, where $|I_{\text{mix}}| > 1$ is fixed in advance. For each $j \in I_{\text{mix}}$, we perform an MCMC run to approximate the corresponding posterior distribution and draw samples $\{\theta_j^{(s)}\}_{s=1}^S \sim p(\theta \mid \tau^{(j)})$. This yields a total of $|I_{\text{mix}}| \cdot S$ posterior samples for constructing the mixture distribution.

We then define the mixture proposal as
\begin{align}
    p_{\rm mix}(\theta \mid \{ \tau^{(j)}\}) = \frac{1}{|I_{\text{mix}}|} \sum_{j \in I_{\text{mix}}} p(\theta\mid\tau^{(j)}) .
\end{align}
To draw samples from $p_{\rm mix}(\theta \mid \{\tau^{(j)} \} )$, we uniformly resample from the combined set of all samples $\{\theta_j^{(s)}\}_{s=1}^S$ for $j \in I_\text{mix}$. We denote the resulting resampled set by $\{\theta_*^{(s)}\}_{s=1}^S$.

This mixture distribution serves as the proposal distribution in importance sampling, enabling us to compute the importance ratios (cf. Equation~(\ref{eq:importance_ratios})). Applying Bayes’ theorem to compute the ratios for each component $i \in I$ yields:

\begin{align}
    r_i(\theta^{(s)}_*) &= \frac{p(\theta^{(s)}_*\mid\tau^{(i)})}{p_{\rm mix}(\theta_*^{(s)} \mid \{\tau^{(j)} \} )} \\ &= \frac{p(\theta^{(s)}_*\mid\tau^{(i)})}{ \frac{1}{\mid I_{\text{mix}}\mid} \sum_{j \in I_{\text{mix}}} p(\theta_*^{(s)}\mid\tau^{(j)})} \\
    &= \frac{p(\tau^{(i)}\mid\theta^{(s)}_*) p(\theta^{(s)}_*)}{p(\tau^{(i)}) \frac{1}{\mid I_{\text{mix}}\mid} \sum_{j \in I_{\text{mix}}} \frac{p(\tau^{(j)}\mid\theta^{(s)}_*) p(\theta^{(s)}_*)}{p(\tau^{(j)})}} \\
    &= \frac{p(\tau^{(i)}\mid\theta^{(s)}_*)}{p(\tau^{(i)}) \frac{1}{|I_{\text{mix}}|} \sum_{j \in I_{\text{mix}}} \frac{p(\tau^{(j)}\mid\theta^{(s)}_*)}{p(\tau^{(j)})}} \\
    &\propto \frac{p(\tau^{(i)}\mid\theta^{(s)}_*)}{\sum_{j \in I_{\text{mix}}} \frac{p(\tau^{(j)}\mid\theta^{(s)}_*)}{p(\tau^{(j)})}}
    \label{eq:importance_ratio_mixture}\, .
\end{align}
In the standard setting of self-normalized importance sampling, the normalizing constants --- here the marginal likelihoods $p(\tau^{(i)})$ --- can be ignored since the self-normalization automatically corrects for them. However, in the case of a mixture proposal, the constants of the mixture components cannot be factored out of the summation in the denominator, and thus self-normalization is no longer directly applicable.
Therefore, to evaluate Equation~\eqref{eq:importance_ratio_mixture}, we require either the exact values or relative magnitudes of the marginal likelihoods $p(\tau^{(j)})$ for all $j \in I_\text{mix}$. For the purpose of estimating $p(\tau^{(j)})$, we use bridge sampling \citep{meng_simulating_1996}, a method for estimating normalizing constants in complex posterior distributions \citep{gronau_bridgesampling_2021}. With these estimates, the mixture-based importance weights can be accurately computed, even though it adds computational complexity.

In our approach, we employ both single and mixture proposal distributions for PSIS, whereas IWMM is applied only with single proposals. Preliminary investigations (not shown here) indicated that using these mixtures within IWMM performs worse than using just a single posterior $p(\theta \mid \tau^*)$ as proposal. We attribute this to the nature of the IWMM method, which relies on affine transformations to shift and scale the proposal distribution to match the moments of the weight distribution \citep{paananen_implicitly_2021}. A single posterior distribution is more likely to share structural similarities --- differing primarily in location and scale --- with another posterior (the target) than a multi-modal mixture distribution. Based on these findings, we opted not to further consider mixture proposals when applying IWMM.

\subsection{Iterative Method \label{sec:iterative_method}}
In this subsection, we integrate the methodological components introduced above to construct an iterative algorithm. The goal is to efficiently approximate the posterior distributions $p(\theta \mid \tau^{(i)})$ for all values $\{\tau^{(i)} \}_{i=1}^m$, and to enable the computation of the Monte Carlo estimate defined in Equation~(\ref{eq:mc_estimate}).
To achieve this, representative values are selected based on a selection strategy (see Section \ref{sec:method_components}).
Algorithm \ref{algo:method} displays the proposed method when selecting a single value $\tau^*$ to calculate the proposal distribution. It is also possible to use a mixture of multiple draws for a proposal as explained above. The algorithm for this special case is displayed in the Appendix \ref{sec:ap:algo_mix} as Algorithm \ref{algo:method_mix}. Either way, we access the proposal through its set of draws $\{\theta_*^{(s)} \}_{s=1}^S$.

For all remaining $\tau^{(i)}$, PSIS is applied, treating ${p(\theta\mid\tau^{(i)})}$ as the target distribution and ${p(\theta\mid\tau^*)}$ as the proposal distribution. If the diagnostic metric $\hat{k}_i$ is below the threshold $k_\text{threshold} $, PSIS is considered successful, and importance resampling is performed to obtain posterior draws from the target distribution. However, if PSIS fails, indicated by $\hat{k}_i \geq k_\text{threshold} $, IWMM is applied.  

If IWMM succeeds, it returns updated draws and importance weights, enabling importance resampling to approximate the desired posterior distribution. If both PSIS and IWMM fail, a new proposal distribution must be selected that better approximates the missing posterior distributions. This results in an iterative algorithm, as outlined in Algorithm \ref{algo:method}.

\begin{algorithm*}
\caption{Iterative algorithm for efficiently calculating the marginalized posterior via the Monte Carlo Estimate using PSIS and IWMM.\label{algo:method}}
\begin{algorithmic}[1]  
\Require Variable of interest $\theta \sim p(\theta \mid C)$, underlying variable $\tau \sim p(\tau \mid C)$
\State Draw $m$ samples $\tau^{(1)}, \dots, \tau^{(m)} \sim p(\tau \mid C)$
\State Set index set $I \gets \{1,\dots, m\}$
\While{$|I| > 0$\label{algo:row:while_loop}}
    \State Select representative value $\tau^* \in\{ \tau^{(i)}\}_{i \in I}$ and run MCMC to access the posterior distribution 

    through samples $\{\theta^{(s)}_*\}_{s=1}^S \sim p(\theta \mid \tau^*)$. \label{algo:row:selection}
    \State Update index set $I$ such that $\tau^* \notin \{\tau^{(i)}\}_{i \in I} $.
    \For{$i \in I$}
        \State Run PSIS with target $p(\theta\mid\tau^{(i)})$, proposal $p(\theta\mid\tau^*)$, obtain importance weights and metric $\hat{k}_i$ \label{algo:row:PSIS}
        \If{$\hat{k}_i < 0.7$}
            \State Use calculated importance weights and draws $\{\theta^{(s)}_*\}_{s=1}^S \sim p(\theta\mid\tau^*)$ for importance 
            
            \hspace{3em} resampling and save the resampled draws as posterior draws corresponding to $p(\theta\mid\tau^{(i)})$.
            \State Remove $i$ from $I$: $I \gets I \setminus \{i\}$
        \Else \label{algo:row:IWMM_start}
            \State Run IWMM with target $p(\theta\mid\tau^{(i)})$ and proposal $p(\theta\mid\tau^*)$ 
            
            \hspace{3.5em}and gain importance weights and metric $\hat{k}_{\text{MM},i}$.
            \If{$\hat{k}_{\text{MM}, i} < 0.7$}
                \State Use calculated importance weights and transformed draws $\{\breve{\theta}^{(s)}_*\}_{s=1}^S$ for importance
                
                \hspace{4.5em} resampling and save the resampled draws as posterior draws corr. to ${p(\theta\mid\tau^{(i)})}$.
                \State Remove $i$ from $I$: $I \gets I \setminus \{i\}$
            \EndIf
        \EndIf \label{algo:row:IWMM_end}
    \EndFor
\EndWhile
\State Calculate the marginalized posterior via Monte Carlo Estimate $p(\theta \mid C) \approx \frac{1}{m} \sum_{i=1}^m p\left(\theta \mid \tau^{(i)}\right)$.
\end{algorithmic}
\end{algorithm*}

Algorithm \ref{algo:method} provides accurate posterior draws through MCMC or resampled posterior draws through PSIS or IWMM for all $\{\tau^{(i)} \}_{i=1}^m$, corresponding to the desired distribution ${p(\theta \mid \tau^{(i)})}$. This allows the calculation of the Monte Carlo estimate from Equation (\ref{eq:mc_estimate}), enabling us to approximate the marginalized posterior distribution $p(\theta \mid C)$. Since each posterior approximation used in the calculation is either based on computationally expensive MCMC or accepted only when the corresponding PSIS diagnostic, $\hat{k}$ or $\hat{k}_{\rm MM}$ is below the threshold, the procedure includes a diagnostic check intended to retain reliable approximations of the individual posterior distributions. The algorithm improves efficiency compared to running MCMC for all $m$ values separately, as each iteration (cycles of the while-loop in row \ref{algo:row:while_loop}) requires only one MCMC run for the proposal distribution, reducing the number of MCMC runs significantly. The improvement in efficiency is tested through simulation studies in Section \ref{sec:experiments}.

The presented method provides an efficient solution for general two-step modeling workflows, potentially applicable beyond the two settings considered here. In this paper, we illustrate its use with two specific applications: a missing data problem and surrogate modeling. For these settings, additional notation is required, which is explained below.

\subsection{Special Case: Missing Data Problem \label{sec:special_case:MI}}

We now present a specific application to demonstrate our method: a regression task on a dataset with missing values \citep{fahrmeir_regression_2013}. The missing data introduces uncertainty, which we aim to propagate to the posterior of the model parameters. In this context, the values $\tau^{(1)}, \dots, \tau^{(m)}$ represent different imputed versions of the dataset, later denoted as $D^{(1)}, \dots, D^{(m)}$. The formal problem definition follows.

\textbf{Problem Definition}\quad
In a regression modeling context, we use a dataset $D$ consisting of $N$ observations of a target variable $y$, along with corresponding observations of $p$ covariates ${X = (x_1, \dots, x_p)}$. With a regression model we aim to describe the relationship between the target variable and the covariates.
The model is defined by a set of parameters $\theta$, their prior distribution $p(\theta)$, and a likelihood function $p(y \mid X, \theta)$, which specifies the assumed response distribution. The central quantity of interest is the posterior distribution $p(\theta \mid D)$ of the parameters given the data \citep{gelman_bayesian_2013}.

\begin{figure*}
    \centering
    \includegraphics[width=\linewidth]{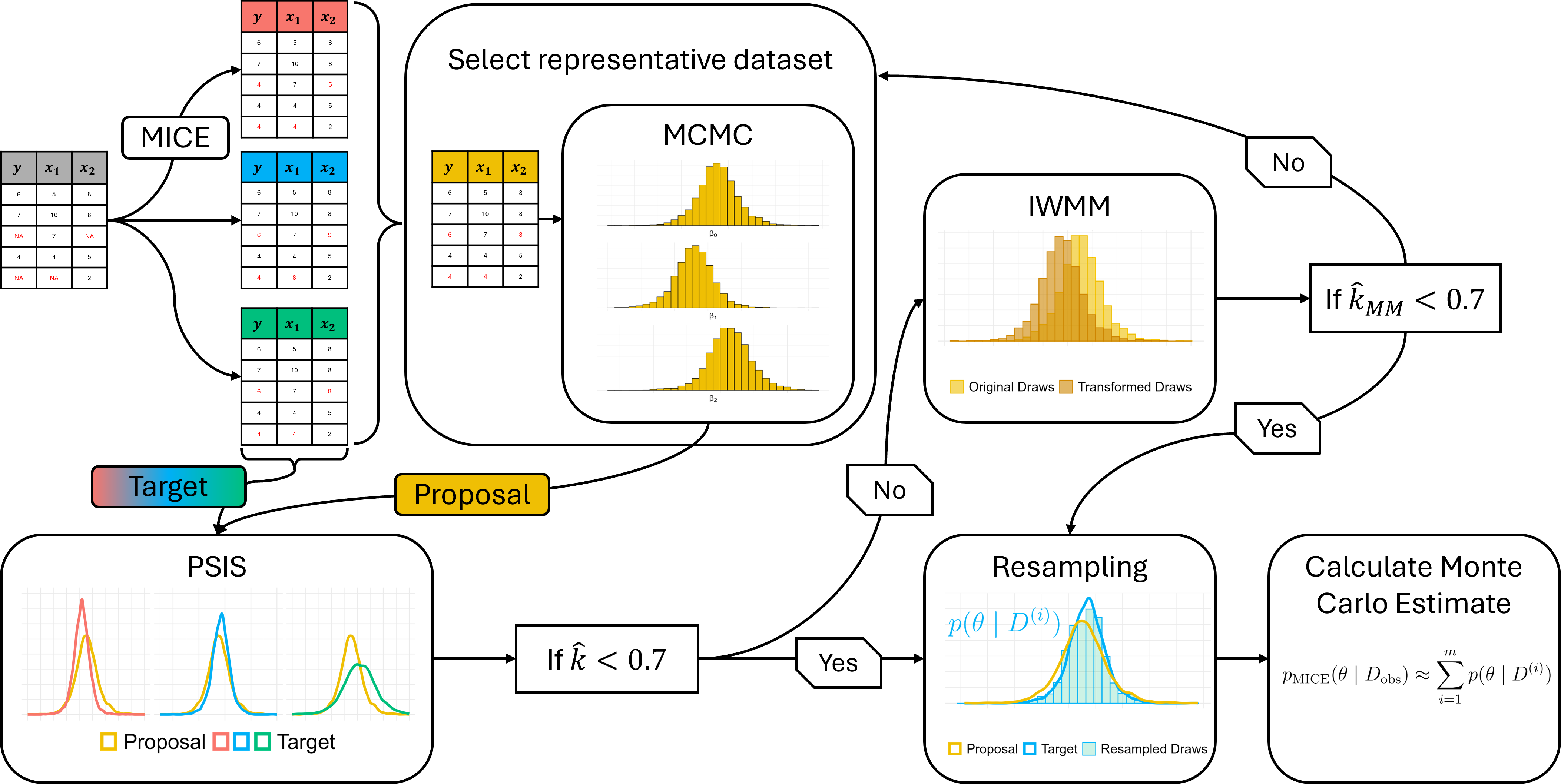}
    \caption{Iterative Algorithm (cf. Algorithm \ref{algo:method}) for the special case of the missing data problem. Multiple datasets are generated by the MICE algorithm, representing the uncertainty induced by the missing values in the original dataset. A representative dataset is selected and the posterior draws accessed through MCMC. This posterior is used to approximate the posteriors of the resulting imputed datasets with PSIS and IWMM, until all posteriors can be approximated using resampling and the Monte Carlo estimate can be calculated.}
    \label{fig:algo_MI}
\end{figure*}

We now consider the setting in which the observed dataset $D_{\rm obs} = (y_{\rm obs}, X_{\rm obs})$ contains missing values, potentially affecting both the target variable and the predictors. While there are straightforward approaches to handling missing data --- such as listwise or pairwise deletion, and mean imputation \citep{buuren_flexible_2018} --- these methods fail to propagate the uncertainty introduced by the missing values into the parameter estimates.
A more principled approach that allows for uncertainty propagation is Multivariate Imputation by Chained Equations (MICE) \citep{van_buuren_flexible_1999}. MICE is an iterative algorithm that imputes the missing values in the dataset by generating $m$ completed versions of the data, each containing potentially different imputed values at the locations of the missing entries \citep{white_multiple_2011}. These multiple imputations help capture the uncertainty associated with the missing values in the original dataset $D_{\rm obs}$.
We define the $m$ imputed datasets as $D^{(1)}, \dots, D^{(m)}$, with observations in each dataset $D^{(i)}$ indexed by the superscript $(i)$ and denote the imputation process performed by MICE via the distribution $p_{\text{MICE}}(D \mid D_{\rm obs})$.
Our goal is to characterize the marginalized posterior distribution of the model parameters $\theta$, which is obtained by integrating over the uncertainty in the imputed datasets:
\begin{align}
    p_\text{MICE}(\theta\mid D_{\rm obs} ) =  \int p&(\theta \mid D) \, p_\text{MICE}(D \mid D_{\rm obs}) \, dD \\ \overset{\text{Monte Carlo}}{\approx}& \frac{1}{m} \sum_{i=1}^m p(\theta \mid D^{(i)}) .
\end{align}
For this problem setting, our method presented in Section \ref{sec:iterative_method} can be applied, with the context $C$ being the observed dataset $D_{\rm obs}$. A structural visualization of the iterative method (cf. Algorithm \ref{algo:method}) is illustrated in Figure~\ref{fig:algo_MI} for the special case of working with missing data.

\textbf{Selection of Representative Datasets}\quad
To run the iterative method (cf. Section \ref{sec:iterative_method}), we first need to select one or more representative datasets. These datasets should be chosen such that their corresponding posterior distributions, $p(\theta \mid D^{(i)})$, adequately represent the overall posterior distribution space. If only a single dataset $D^*$ is selected, it is desirable for its posterior distribution to be centrally located within this space. In the case of multiple selected datasets, their posterior distributions should collectively cover the space appropriately.  

Since we do not have direct access to the posterior distributions corresponding to the datasets without running computationally expensive methods such as MCMC, we assume that if two datasets, $D^{(i)}$ and $D^{(j)}$, are highly similar, their corresponding posterior distributions, $p(\theta \mid D^{(i)})$ and $p(\theta \mid D^{(j)})$, will also be similar. Based on this assumption, we can select representative datasets using similarity measures in the data space. The posterior distributions associated with these representative datasets can then be used as proposal distributions in PSIS and IWMM.

To quantify the pairwise similarity between datasets, an appropriate distance measure must be chosen \citep{stolte_methods_2024}. One possible approach is the Friedman-Rafsky test \citep{friedman_multivariate_1979}, which constructs a pooled dataset from $D^{(i)}$ and $D^{(j)}$ and forms a complete graph where each sample is represented as a node. The edge weights between nodes are computed based on Euclidean distances between samples. A minimum spanning tree \citep{cormen_introduction_2009} is then constructed, and a distance measure is derived by counting the number of edges in the tree that connect observations from different datasets.  

By computing pairwise distances for all $m$ imputed datasets, we form a distance matrix. We use this matrix with a clustering algorithm such as $k$-medoids \citep{kaufman_finding_1990} to identify a centered dataset or, for $k > 1$, a subset of representative datasets that are well-distributed across the dataset space. For simplicity, we consider the selection of a single dataset $D^*$ in the following discussion. 
When using a mixture proposal as described in Section~\ref{sec:mixture}, we select multiple datasets with the $k$-medoids algorithm for $k>1$.

\textbf{Calculation of Importance Ratios}\quad
The specific case of handling missing data offers additional advantages that allow our algorithm to operate even more efficiently. In particular, the computation of importance ratios for PSIS and IWMM can be further optimized in this setting --- an improvement not necessarily feasible in the more general iterative approach described in Section~\ref{sec:iterative_method}. This efficiency gain stems from the structural similarity among the imputed datasets $\{D^{(i)}\}_{i \in I}$, which differ only in the entries corresponding to originally missing values. We denote by $J^* \subset \{1, \dots, N\}$ the index set of observations where these differences occur.

Let $D^*$ be the dataset selected as the representative from the set $\{D^{(i)}\}_{i \in I}$. Using MCMC, we draw samples $\{\theta_*^{(s)}\}_{s=1}^S \sim p(\theta \mid D^*)$ to approximate the posterior given $D^*$. These samples can then be used to compute importance ratios for the remaining datasets $D^{(i)}, i \in I$, as follows:
\begin{align}
    r_i(\theta_*^{(s)}) &= \frac{p(\theta_*^{(s)}\mid D^{(i)})}{p(\theta_*^{(s)}\mid D^*)} = \frac{p(\theta_*^{(s)}\mid y^{(i)}, X^{(i)})}{p(\theta_*^{(s)}\mid y^*, X^*)}  \\ &\propto \frac{p(y^{(i)}\mid \theta_*^{(s)}, X^{(i)})p(\theta_*^{(s)})}{p(y^*\mid\theta_*^{(s)}, X^*)p(\theta_*^{(s)})} \\ &= \frac{p(y^{(i)}\mid  \theta_*^{(s)}, X^{(i)})}{p(y^*\mid \theta_*^{(s)}, X^*)} \, , s=1, \dots, S.
\end{align}
We represent the importance ratios using a pointwise likelihood formulation, leveraging the conditional independence of observations, which allows the likelihood to factorize \citep{burkner_efficient_2020}. Given that the datasets differ only in the subset of rows indexed by $J^*$, we obtain
\begin{align}
    r_i(\theta_*^{(s)}) &= \frac{\prod_{n=1}^N p(y_n^{(i)}\mid \theta_*^{(s)}, X^{(i)})}{\prod_{n=1}^N p(y_n^*\mid \theta_*^{(s)}, X^*)} \\ &= \frac{\prod_{n\in J^*} p(y_n^{(i)}\mid \theta_*^{(s)}, X^{(i)})}{\prod_{n\in J^*} p(y_n^*\mid \theta_*^{(s)}, X^*)} \, , \label{eq:MI_ratio_kuerzen}
\end{align}
$s=1, \dots, S $, because $p(y_n^{(i)} \mid \theta_*^{(s)}, X^{(i)}) = {p(y_n^* \mid \theta_*^{(s)}, X^*)}$ for all $n\notin J^*$. This reduction allows us to evaluate only the likelihood terms that correspond to differing rows, decreasing the number of pointwise likelihood evaluations by a factor of $|J^*| / N$, where $|J^*|$ corresponds to the number of rows with missing values.
This optimization significantly reduces the computational effort required to compute the importance ratios for PSIS and IWMM, thereby making the overall workflow more efficient.

\subsection{Special Case: Surrogate Modeling}\label{sec:special_case:surrogate}
The second application of our method is in the context of surrogate modeling. Simulators are essential tools for studying complex natural systems, as they allow researchers to replicate, manipulate, and analyze theoretical models of real-world phenomena under controlled and reproducible conditions \citep{winsberg_science_2010, parker_does_2009}. However, these simulators are often computationally intensive and time-consuming to evaluate \citep{peck_simulation_2004}. To alleviate this burden, surrogate models are employed as efficient approximations of the original simulators. These surrogate models aim to replicate the behavior of the simulator while being significantly faster to evaluate and simpler to work with \citep{zhu_bayesian_2018, lavin_simulation_2022}.

An important challenge in this setting is the propagation of uncertainty arising both from the data simulated from a potentially stochastic simulator and from the approximation error of the surrogate model \citep{reiser_uncertainty_2025}. Properly accounting for this uncertainty is crucial for reliable inference, making surrogate modeling a compelling application of our proposed method.

\textbf{Problem Definition}\quad
We begin by considering a complex simulator $\mathcal{M}$, which may represent, for example, a physical or biological system. Due to its complexity, evaluating $\mathcal{M}$ can be computationally prohibitive. Given input parameters $\theta_T$, the simulator produces an output \citep{reiser_uncertainty_2025}:
\begin{align}
    y_T = \mathcal{M}(\theta_T) \, .
\end{align}
The output $y_T$ is assumed to be generated from the true distribution $p(y_T \mid \theta_T)$, governed by the simulation model. Due to the simulator's complexity, however, this likelihood is typically intractable and computationally difficult to evaluate.

To approximate this process, we introduce a surrogate model $\widetilde{\mathcal{M}}$ such that $\widetilde{\mathcal{M}} \approx \mathcal{M}$. The surrogate model is designed to be computationally more tractable while maintaining a reasonable approximation to the true simulator. It is parameterized by trainable parameters $\tau$, leading to the following form \citep{reiser_uncertainty_2025}:
\begin{align}
    \tilde{y}_T = \widetilde{\mathcal{M}}(\theta_T, \tau) \, .
\end{align}

\textbf{Surrogate Training Step}\quad 
In this step, we train the surrogate model to approximate the behavior of the simulator. To generate training data, we evaluate the simulator using a set of input parameters $\theta_T^{(n)}$, resulting in corresponding outputs $y_T^{(n)}$. This yields a training dataset denoted by ${D_T =\{\theta_T^{(n)}, y_T^{(n)}\}_{n=1}^{N_T}}$. 

Since simulator evaluations are often computationally expensive, the size of the training dataset $D_T$ is typically small. Despite this constraint, we aim to estimate the surrogate parameters $\tau$ such that the surrogate model accurately captures the input-output relationship of the simulator.

To this end, we adopt a Bayesian approach to infer the surrogate parameters \citep{gelman_bayesian_2013}. Unlike the simulator, for which the true data-generating likelihood $p(y_T \mid \theta_T)$ is analytically intractable, the surrogate model provides a tractable likelihood $p(y_T\mid \theta_T, \tau)$. Using this surrogate likelihood and a prior distribution $p(\tau)$ on the surrogate parameters, we construct the posterior distribution:
\begin{align}
    p(\tau\mid D_T) \propto \prod_{n=1}^{N_T} p(y_T^{(n)}\mid \theta_T^{(n)}, \tau) \, p(\tau) \,.
\end{align}
We approximate this posterior using sampling-based methods such as MCMC \citep{robert_monte_2004}, obtaining a set of samples $\{\tau^{(i)}\}_{i=1}^m \sim p(\tau \mid D_T)$. These samples capture the uncertainty inherent in fitting the surrogate model to the sparse training data generated by the simulator. In the next step, this uncertainty is propagated to the inference task involving new, unseen data.

\textbf{Inference Step}\quad
Next, we consider a new set of observations --- typically real-world measurements --- denoted by ${D_I = \{y_I^{(n)} \}_{n=1}^{N_I}}$. We assume that the (implicit) data-generating process underlying these observations is well captured by the simulator $\mathcal{M}$. In this context, the measurement data $D_I$ is thought to be generated by the simulator using some unknown input parameter $\theta_I$:
\begin{align}
    y_I = \mathcal{M}(\theta_I) \, .
\end{align}

Unlike in the simulation setup, we do not know the input parameters $\theta_I$ that generated the outputs. Our goal is to infer $\theta_I$ given the observed outputs $D_I$.

Since the simulator $\mathcal{M}$ does not have a tractable likelihood function, the posterior distribution of $\theta_I$ cannot be computed directly using Bayes' theorem. Instead, we rely on the surrogate model, specifically using the posterior samples of its parameters $\{ \tau^{(i)} \}_{i=1}^m$ obtained during the training phase.  
The marginalized posterior distribution can be expressed as:
\begin{align}
    p(\theta_I\mid  D_T, D_I) &= \int p(\theta_I\mid \tau, D_I) \, p(\tau\mid D_T) \, d\tau \\ &\overset{\rm MC}{\approx} \frac{1}{m} \sum_{i=1}^m p(\theta_I\mid \tau^{(i)}, D_I) \,.
\end{align}
To compute this Monte Carlo estimate, we must evaluate $p(\theta_I\mid \tau^{(i)}, D_I)$ for each surrogate parameter sample $\tau^{(i)}$. This inference task aligns with the framework introduced in our iterative method (see Section \ref{sec:iterative_method}), which provides an efficient approach to solving this class of problems.

\textbf{Selection of Representative Draws}\quad
To apply our iterative method, we require a selection strategy to identify representative draws from the set $\{ \tau^{(i)} \}_{i \in I}$. As discussed in Section~\ref{sec:method_components}, this strategy depends on the specific problem setting. While one approach was outlined for the case of missing data in Section \ref{sec:special_case:MI}, a different strategy is needed in the context of surrogate modeling.

The selected draws should be representative in the sense that their corresponding posterior distributions collectively cover the diversity of possible posterior shapes --- or, in the case of selecting a single draw, that the draw is central in terms of its posterior characteristics.

A straightforward approach is to apply clustering algorithms such as $k$-means~\citep{macqueen_methods_1967} or $k$-medoids~\citep{kaufman_finding_1990} in the space of the draws $\tau$. However, we cannot be sure that the similarity between surrogate parameter draws $\tau^{(i)}$ and $\tau^{(j)}$ does reliably reflect the similarity of their respective posterior distributions, $p(\theta_I \mid \tau^{(i)}, D_I)$ and $p(\theta_I \mid \tau^{(j)}, D_I)$, due to the non-linear transformation from the surrogate parameter space into the posterior space.

Instead, we aim to infer information about the location and structure of the posteriors ${p(\theta_I \mid \tau^{(i)}, D_I)}$ without resorting to expensive methods such as MCMC or requiring knowledge of the true parameter $\theta_I$. To this end, we draw samples $\{\theta_I^{(j)}\}_{j=1}^J\sim p(\theta_I)$ from the prior. For each draw of surrogate parameters $\tau^{(i)}$, we compute the surrogate log-likelihood of the observed data $D_I$ as
\begin{equation}
    L_i (\theta_I^{(j)}) = \sum_{n=1}^{N_I} \log p(y_I^{(n)}\mid \theta_I^{(j)}, \tau^{(i)}),
\end{equation}
and average these log-likelihoods over all prior samples:
\begin{equation}
    \bar{L}_i = \frac{1}{J} \sum_{j=1}^J L_i(\theta_I^{(j)}).
\end{equation}
Ranking the surrogate parameter draws $\tau^{(i)}$ by their corresponding $\bar{L}_i$ values provides insight into the relative positions of the associated posterior distributions.  
This ranking allows us to select representative draws based on quantiles of the sorted $\bar{L}_i$ values.  
For instance, when selecting $N_\text{mix} = 5$ representative draws for a mixture distribution, we take the draws corresponding to the minimum and maximum $\bar{L}_i$ values, as well as the 25\textsuperscript{th}, 50\textsuperscript{th}, and 75\textsuperscript{th} percentiles.

\begin{figure}
    \centering
    \includegraphics[width=\linewidth]{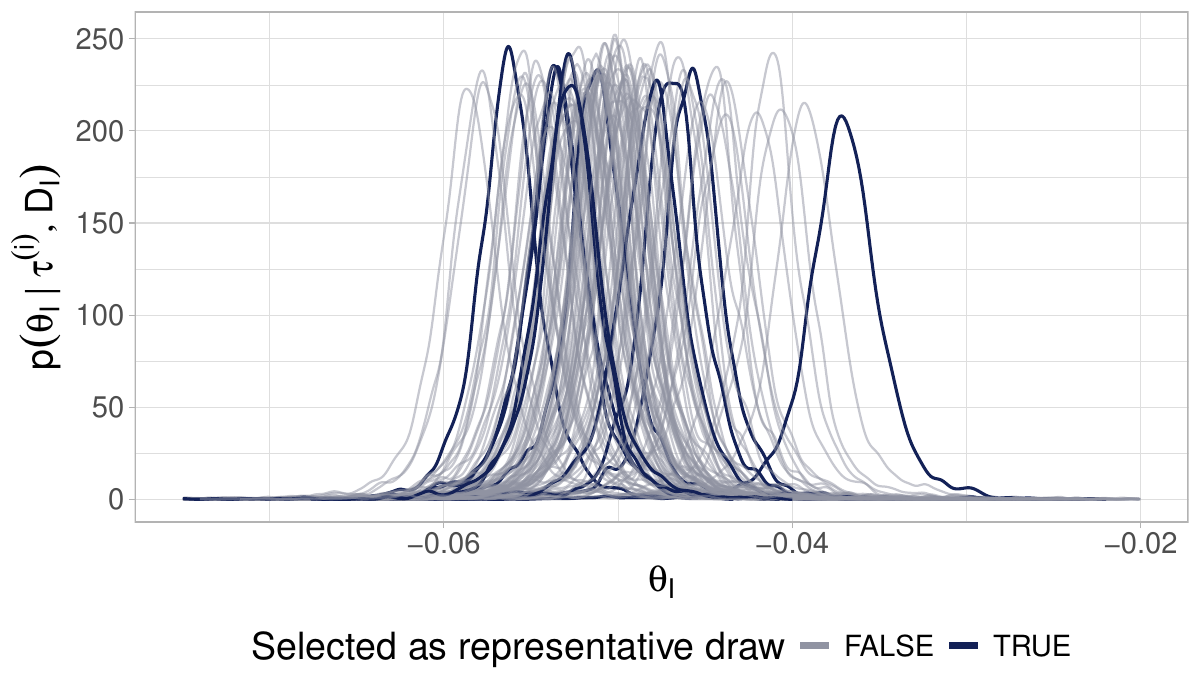}
\caption{Posterior distributions $p(\theta_I \mid \tau^{(i)}, D_I)$ for $i=1,\dots,100$ from the logistic-surrogate simulation study in Section~\ref{sec:surrogates_sim_study}. 
The $N_\mathrm{mix}=10$ distributions highlighted in blue were selected by the proposed method based on log-likelihoods evaluated over prior samples.}
    \label{fig:surrogate_llmethod_posteriors}
\end{figure}

Figure~\ref{fig:surrogate_llmethod_posteriors} provides an exemplary visualization of the selection results. It shows the posterior distributions $p(\theta_I \mid \tau^{(i)}, D_I)$ for all surrogate realizations $\tau^{(i)}$ in the logistic-surrogate simulation study (Section~\ref{sec:surrogates_sim_study}). 
The $N_\mathrm{mix}=10$ distributions highlighted in blue are those selected by the proposed method. The selected posteriors are well distributed across the range of plausible values of $\theta_I$, indicating broader coverage than would typically be obtained from a random subset.

\section{Experiments \label{sec:experiments}}

To showcase our proposed method for propagating uncertainty in practice, we conduct multiple numerical experiments for both the missing value problem and the surrogate modeling.
In addition, we assess its performance in a real-world case study using a dataset of taxi trips in New York City that contains missing values. All code and material can be found on GitHub\footnote{\url{https://github.com/sjedhoff/efficient-2step-uncertainty-paper}}.

\subsection{Multiple Imputation -- Simulation Study \label{sec:MI_sim_study}}

\subsubsection{Simulation Setup}
For evaluating the performance of the proposed method under different settings, we simulate a variety of datasets. A dataset $D$ was generated with $N \in \{ 10,100\}$ observations for a single target variable $y$ and $p$ covariates, where $p \in \{2,5\}$ for $N=10$ and $p \in \{10,30 \}$ for $N=100$. The datasets were generated based on the structure of a linear regression model:
\begin{align}
    y = \beta_0 + \beta_1 x_1 + \beta_2 x_2 + \dots + \beta_p x_p + \epsilon,
\end{align}
$\epsilon \sim \mathcal{N}(0,1)$. The model parameters ${\mathbf{\beta} = {(\beta_0, \beta_1, \dots, \beta_p)}}$ were drawn from a $\mathcal{N}(0,1)$ distribution, while the covariates $x_i$ were generated with a mean of 0 and standard deviation sampled from a Gamma distribution with shape and rate parameters set to 10. The datasets were constructed using the \texttt{R}-package \texttt{simstudy} \citep{goldfeld_simstudy_2024}, enabling the generation of correlated data with a correlation coefficient of 0.3 between variables. For each dataset size, 20 datasets were created. Missing values were then introduced in $\lceil \pi \cdot N \rceil$ rows, with $\pi \in \{0.05,0.15,0.30\}$. Within each affected row, the target variable $y$ was always missing, along with $\lfloor \frac{p}{2} \rfloor$ randomly selected covariates. 

Each dataset was subsequently imputed ${m=100}$ times using the MICE algorithm, implemented in the \texttt{R}-package \texttt{mice} \citep{buuren_flexible_2018, buuren_mice_2011}.
Choosing the number of imputations $m$ is nontrivial and depends on several factors such as the fraction of missing information, the desired stability of the resulting inference, and practical constraints in the application. Further guidance on selecting $m$ is provided in \cite{graham_how_2007, von_hippel_how_2020}. In our setting, we examined how the posterior distributions depend on $m$ and on the number of samples drawn per imputed dataset $S$. Under the gold-standard approach --- running MCMC separately for each imputed dataset --- we observed no visible differences between the resulting posterior distributions for $m=10,S=400$; $m=100, S=40$ and $m=1000,S=4$ in our case (see Figure~\ref{fig:posteriors_vary_mS} in Appendix~\ref{sec:ap:MI_sim_results}).

The models were fitted on the imputed datasets via MCMC with the \texttt{R}-package \texttt{brms} \citep{burkner_brms_2017, burkner_brms_2024, stan_development_team_stan_2024} with four chains for 2000 iterations each (including 1000 warm-up and 1000 sampling iterations). Two different prior specifications were tested: the default noninformative (flat) priors in \texttt{brms} and the regularized Horseshoe prior \citep{carvalho_horseshoe_2010, piironen_hyperprior_2016}, which is employed for regularization and model selection \citep{carvalho_handling_2009}. 

The simulation study evaluates four different methods.  
The first is a brute-force approach that fits each of the $m = 100$ imputed datasets individually using \textit{MCMC}. This computationally intensive method is implemented using \texttt{brms}.
The second method employs the iterative algorithm described in Section \ref{sec:iterative_method}, but omits the IWMM step. Specifically, after executing the PSIS step (Line \ref{algo:row:PSIS} in Algorithm \ref{algo:method}) and verifying its success for the given target distribution, the algorithm skips the IWMM step (Lines \ref{algo:row:IWMM_start} to \ref{algo:row:IWMM_end}) and proceeds with a new proposal distribution. This variant is referred to as \textit{PSIS:single}.
The third method, termed \textit{PSIS:mixture}, also relies solely on PSIS but utilizes a mixture distribution with $|I_{\rm mix} | = 5$, as outlined in Algorithm \ref{algo:method_mix} in Appendix \ref{sec:ap:algo_mix}.
The fourth method, introduced in Section \ref{sec:iterative_method}, integrates both PSIS and IWMM --- jointly referred to as \textit{PSIS+IWMM} in the following. Preliminary simulations revealed that using a mixture proposal distribution within IWMM did not improve performance relative to a single proposal; in fact, it often degraded it. This reduction in performance is likely due to the close similarity in the structure of the target and single proposal distribution, whereas the use of a mixture distribution introduces greater divergence, thereby complicating the transformation step in IWMM. Therefore, no mixture distribution is tested for the \textit{PSIS+IWMM} method.

\textbf{Selection Strategies}\quad
In addition to the different inference methods described above, we also evaluated various strategies for selecting representative datasets to serve as proposal distributions. Three selection strategies were considered. The first, \textit{Random}, simply selects a subset of datasets at random to act as new proposals. The second strategy, \textit{Medoids}, follows the approach described in Section~\ref{sec:special_case:MI}. The third strategy, \textit{Max $\hat{k}$}, chooses datasets with the highest estimated Pareto shape parameter $\hat{k}$ (or $\hat{k}_{\rm MM}$) in each iteration to be used as new proposals. For the initial round, where no $\hat{k}$ estimates are yet available, the \textit{Medoids} strategy was used.

\textbf{Performance Evaluation}\quad
To evaluate the efficiency of the proposed method relative to the brute-force \textit{MCMC} approach, we compare the number of log-likelihood evaluations.  
The accuracy of the resulting marginalized posteriors should be satisfactory, as we only accept importance resampling approximations from PSIS or IWMM when the corresponding shape parameters $\hat{k}$ or $\hat{k}_{\rm MM}$ fall below the threshold $k_{\rm threshold}$.  
This criterion already ensures that only reliable approximations of the target distribution are used \citep{vehtari_pareto_2024}, as exemplified in Figure~\ref{fig:posteriors}. To further validate this, we compute the mean absolute differences between posterior summaries obtained via \textit{MCMC} and via the \textit{PSIS+IWMM} approach, considering the posterior mean, standard deviation, and the 5\% and 95\% quantiles.

\begin{figure*}[ht]
    \centering
    \includegraphics[width=\linewidth]{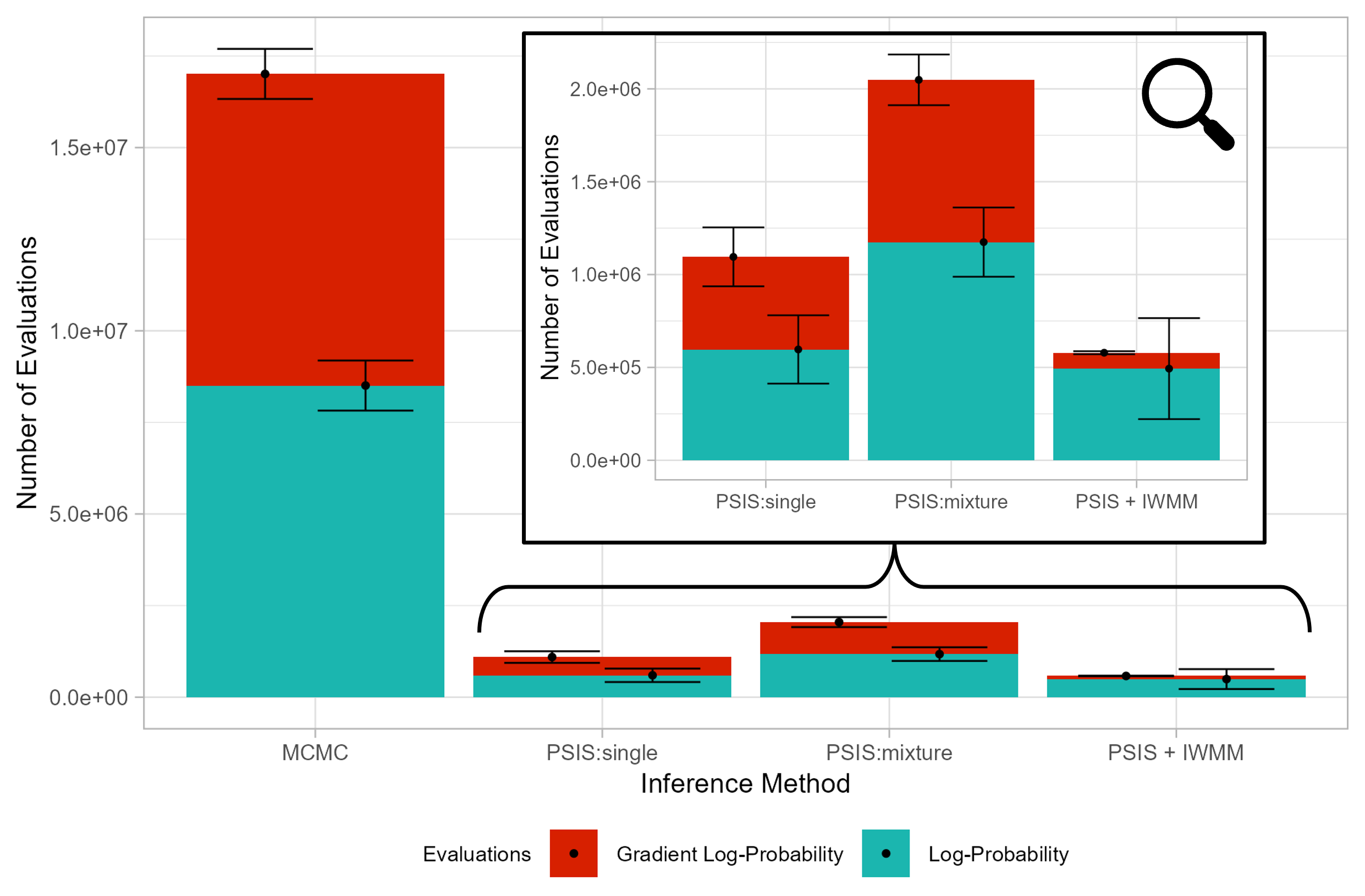}
    \caption{Number of log-probability (blue) and gradient log-probability evaluations (red) required to approximate the posteriors of $m = 100$ imputed datasets ($N = 100$, $p = 10$) with 15\% of the rows containing missing values in the original data. Representative datasets were selected using the \textit{Medoids} method. Error bars indicate the standard deviation across repeated runs of the same setup.}
    \label{fig:barplot_evaluation}
\end{figure*}

The models for the PSIS and IWMM proposals are fitted in \texttt{brms} \citep{burkner_brms_2017, burkner_brms_2024} using \texttt{Stan} \citep{stan_development_team_stan_2024}, where Hamiltonian Monte Carlo (HMC) \citep{neal_mcmc_2011}, more specifically the no-U-turn-sampler (NUTS) \citep{homan_no-u-turn_2014}, is employed, as the gold standard for accurate sampling.
The majority of NUTS’s computational cost arises from the repeated evaluation of the log-probability and its gradient, referred to as gradient evaluations. The log-probability evaluations correspond to the total unnormalized log-posterior density, incorporating both the log-prior of the parameters and the log-likelihood of the observed data given the parameters. In contrast, gradient log-probability evaluations refer to computing the derivative of the log-probability function with respect to all model parameters. Evaluating these gradients is crucial for HMC, as it relies on using gradient information to navigate the parameter space effectively. Gradients are computed via automatic differentiation \citep{carpenter_stan_2015}, and typically account for approximately 80\% of the total computational time \citep{zhang_pathnder_2022}. Therefore, the number of gradient evaluations is a key factor when comparing the computational costs of different methods.
In addition, calculating the importance ratios for PSIS and IWMM requires log-likelihood evaluations --- both for the numerator and denominator of the ratios. Due to the structure of the importance ratios, evaluating the log-prior terms separately is unnecessary.
To compare computational effort across methods, we count log-likelihood evaluations from PSIS and IWMM together with the log-probability evaluations used in HMC.  
While this slightly overstates the cost for PSIS and IWMM --- since log-likelihoods are usually cheaper to compute than full log-probabilities --- it offers a consistent and conservative measure of computational demand.

\subsubsection{Results of the Simulation Study}
\begin{figure*}[th]
    \centering
    \includegraphics[width=\linewidth]{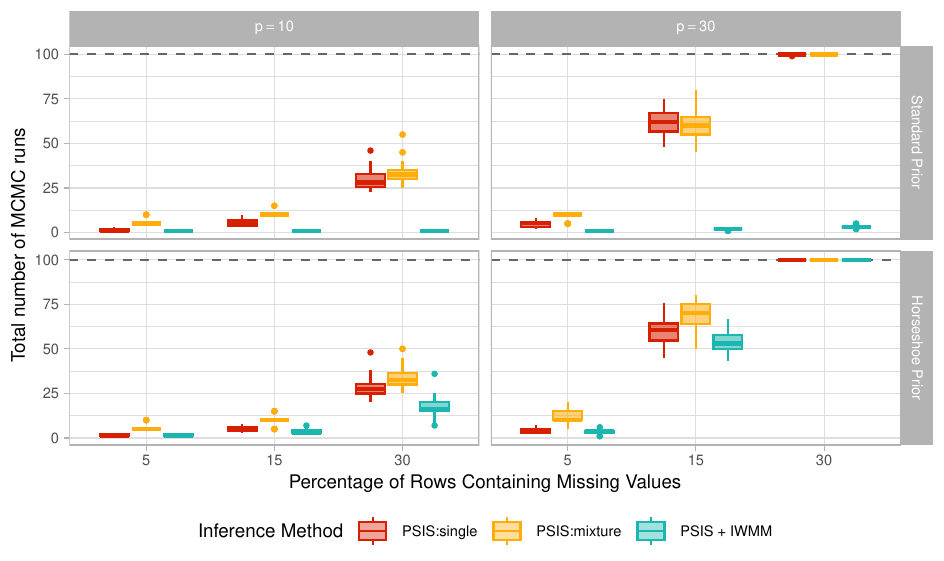}
    \caption{Number of MCMC runs required to approximate the posteriors of $ m = 100$ imputed datasets with $N = 100$ observations, across varying model complexities ($p = 10$ or $p = 30$, columns) and prior specifications (standard \texttt{brms} priors vs. regularized horseshoe, rows). Representative datasets were selected using the \textit{Medoids} procedure. The dashed horizontal reference line at 100 indicates the number of MCMC runs needed when fitting each imputed dataset separately.}
    \label{fig:no_MCMC_n100_both}
\end{figure*}

We now present the findings from the simulation study designed to evaluate the proposed method. We focus on the larger dataset with $N=100$ observations and either $p=10$ or $p=30$ covariates. For the smaller dataset with ${N=10}$ observations, the problem setting is relatively simple, and our proposed method consistently demonstrated reliable performance across all considered scenarios. As a result, we focus on the more challenging and practical case with ${N=100}$ observations. Results for the smaller dataset are provided in Appendix \ref{sec:ap:MI_sim_results}.
Across all experiments, we observed no substantial differences between the various selection strategies for identifying representative datasets. Consequently, we report results using the \textit{Medoids} method only, while additional results for the alternative strategies are included in Appendix~\ref{sec:ap:MI_sim_results}.

Figure \ref{fig:barplot_evaluation} displays the number of log-probability and gradient log-probability evaluations required to approximate the posteriors of all $m=100$ imputed datasets ($N=100$, $p=10$), where in the original dataset 15\% of rows contained missing values.
The results clearly demonstrate that our proposed method substantially reduces the number of required evaluations compared to the brute-force \textit{MCMC} baseline --- both in terms of log-probability and gradient computations. The greatest efficiency is achieved when combining PSIS and IWMM (\textit{PSIS+IWMM}), resulting in the lowest computational cost while preserving inference quality and adequately propagating uncertainty across the imputed datasets.

Figure~\ref{fig:no_MCMC_n100_both} presents the number of MCMC runs required to approximate the posteriors of all ${m = 100}$ imputed datasets. The number of MCMC runs corresponds to the number of distinct proposal distributions needed for successful approximation of all target posterior distributions using our method, compared with the brute-force approach of running \textit{MCMC} for all imputed datasets.
The left panel of the figure shows results for the lower-dimensional case with $p = 10$ predictors. Here, depending on the proportion of missing rows in the original dataset, between 1 and 55 MCMC runs are sufficient --- highlighting the efficiency gains relative to running MCMC for all 100 datasets individually. As expected, the number of required MCMC runs increases with the proportion of missing data. Higher rates of missingness lead to greater variation among the imputed datasets and thus more diverse posterior distributions, making approximation more difficult for \textit{PSIS:single}, \textit{PSIS:mixture} and \textit{PSIS+IWMM}.
In the high-dimensional case with $ p = 30$ predictors (right panel), we observe that PSIS alone --- whether based on a single proposal or a mixture --- fails when the proportion of rows containing missing values is large. Nonetheless, \textit{PSIS+IWMM} continues to offer substantial efficiency gains in most settings using standard priors. However, under the more challenging regularized horseshoe prior and 30\% missing data, \textit{PSIS+IWMM} also struggles to reduce the number of required MCMC runs effectively.

\begin{figure*}
    \centering
    \includegraphics[width=\linewidth]{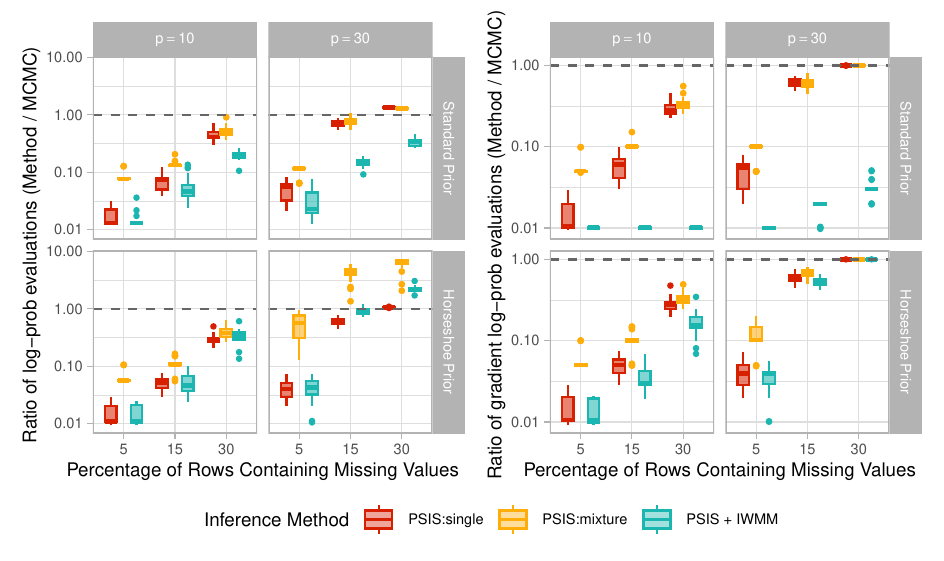}
    \caption{Ratio of number of evaluations (on a log-scale) of the log-probability (\textbf{left}) and of the gradient log-probability (\textbf{right}) compared to running \textit{MCMC} separately necessary to approximate the posteriors of $m = 100$ imputed datasets with size $N=100$ and number of parameters $p$ (in the columns) using either the \texttt{brms} standard priors or the regularized horseshoe prior (in the rows). The representative datasets are selected using the \textit{Medoids} procedure.}
    \label{fig:ratio_evals_n100}
\end{figure*}

While the previous figure focused on the number of MCMC runs, Figure~\ref{fig:ratio_evals_n100} (left) reports the ratio of log-probability evaluations needed by the iterative algorithm relative to the number required with brute-force \textit{MCMC} (baseline).
A ratio close to 1, indicated by the horizontal reference line, implies that the computational effort in terms of log-probability evaluations is comparable to performing 100 separate MCMC runs. For the lower-dimensional setting (left column), the iterative algorithm requires fewer than half as many evaluations, demonstrating substantial computational savings. Results are similar for both the standard priors and the regularized horseshoe prior, though \textit{PSIS+IWMM} performs slightly better with the standard priors.
In contrast, for the higher-dimensional case (right column), \textit{PSIS+IWMM} yields efficiency gains only when standard priors are used. In the most challenging scenario --- 30\% missing data and the horseshoe prior --- the number of log-probability evaluations exceeds that of the brute-force \textit{MCMC} baseline, indicating that the method becomes less efficient in this setting.

A similar pattern is observed when examining the ratio of gradient log-probability evaluations, as shown in Figure~\ref{fig:ratio_evals_n100} (right).  Unlike log-probability evaluations, gradient evaluations are only required during HMC sampling and are not needed for the computation of importance ratios in PSIS or IWMM. Consequently, for the inference methods based on our iterative algorithm (\textit{PSIS:single}, \textit{PSIS:mixture}, and \textit{PSIS+IWMM}), the gradient evaluations originate exclusively from the HMC runs used to generate proposal distributions. This structural difference explains why these methods inherently require fewer or, at most, the same number of gradient evaluations as the brute-force \textit{MCMC} approach.
For the standard priors (top row in Figure~\ref{fig:ratio_evals_n100} right), \textit{PSIS+IWMM} shows even greater improvements in terms of gradient efficiency compared to the gains observed for log-probability evaluations. In the more challenging setting with high dimensionality and the horseshoe prior (bottom right), the performance of \textit{PSIS+IWMM} is comparable to that of \textit{PSIS:single} and \textit{PSIS:mixture}, indicating that in this case, the added complexity of \textit{PSIS+IWMM} does not lead to substantial gains.

Overall, we observe that our proposed method offers substantial improvements in terms of number of log-probability and gradient evaluations --- an indicator of computational cost --- across most settings. For simpler priors, which result in relatively straightforward posterior distributions, the iterative approach combining both PSIS and IWMM significantly outperforms the baseline \textit{MCMC} approach. Even in cases involving more complex posterior shapes, such as those induced by the regularized horseshoe prior, the proposed method continues to outperform alternative inference strategies. The only exception arises in high-dimensional settings (i.e., $p = 30$ covariates) combined with a high proportion of missing data (i.e., 30\% of rows with missing values), where the method does not yield improvements over the brute-force \textit{MCMC} approach.

Our proposed method in the experiments with the regularized horseshoe prior performs worse than in those with the standard priors for the following reason. The stronger shrinkage induces substantially greater variation in the shape of the resulting posteriors across different datasets $D^{(i)}$. In particular, the posterior tail behavior can change markedly (e.g., heavier versus lighter tails), and in some settings the posterior can even become multimodal. As a result, posterior distributions differ much more across $D^{(i)}$, which makes approximation via PSIS and IWMM more challenging and increases the probability of failure, especially when the proposal distribution and the target posterior have fundamentally different shapes.

\begin{figure*}[htb]
    \centering
    \includegraphics[width=\linewidth]{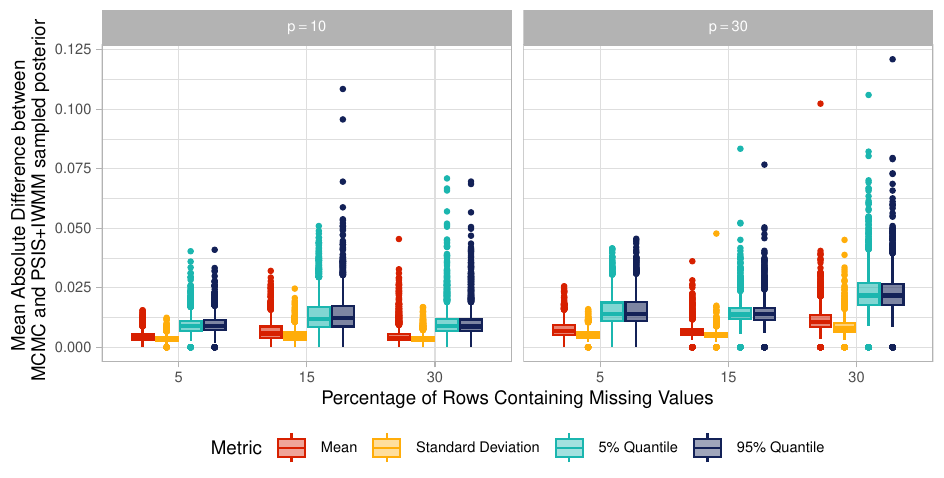}
    \caption{Mean absolute differences between posterior summaries obtained from \textit{MCMC} and from the \textit{PSIS+IWMM} resampled posterior for datasets of size $N=100$. For each missingness level (5\%, 15\%, and 30\% of rows containing missing values), boxplots summarize the distribution across the 20 datasets and their 100 imputed datasets; within each dataset, differences are averaged over the $p$ parameters. Colors indicate the summary metric (posterior mean, standard deviation, 5\% quantile, and 95\% quantile). Results are shown for $p=10$ (left) and $p=30$ (right).}

    \label{fig:mi_difference_posterior_n100}
\end{figure*}

Although our iterative \textit{PSIS+IWMM} approach reduces the number of log-probability and gradient evaluations in most settings, it approximates the posteriors via an additional step by transforming the MCMC samples with PSIS and IWMM, rather than estimating them directly with MCMC. This extra approximation layer can increase Monte Carlo standard errors and, consequently, yield slightly less accurate posterior estimates.
To ensure that the approximation remains reliable, we apply the cutoff $k_{\mathrm{threshold}} = 0.7$ for PSIS and IWMM diagnostics, following the recommendations of \cite{vehtari_pareto_2024}. This threshold is intended to retain only cases where importance sampling provides a sufficiently accurate approximation to the true posterior.

To assess how closely the approximated posterior matches the MCMC baseline, we compare the resampled posteriors to the MCMC posterior using summary statistics. Figure~\ref{fig:mi_difference_posterior_n100} reports absolute differences in the posterior mean, standard deviation, and the 5\% and 95\% quantiles between \textit{PSIS+IWMM} and \textit{MCMC}. The reported values are averaged over the $p$ parameters. Results are shown for datasets with $N=100$, with $p=10$ parameters (left) and $p=30$ parameters (right). Overall, the posterior means are very similar, with average absolute differences below 0.0125. The same is true for the posterior standard deviations. Accurately approximating tail behavior is more challenging, as reflected by larger differences in the 5\% and 95\% quantiles; however, the average absolute differences remain below 0.025.

For a multivariate comparison of the posterior distributions, we additionally use the kernel maximum mean discrepancy (MMD) test \citep{gretton_kernel_2012}. This is a sample-based test with the null hypothesis that two sets of samples are drawn from the same distribution. Under a good approximation, the test should rarely reject the null at significance level $\alpha = 0.05$. For example, for $N=100$, $p=10$, and 30\% missingness, the null is rejected in 4.45\% of cases, which is consistent with the nominal Type~I error rate. In the more challenging setting with $N=10$ and a relatively large number of parameters ($p=5$), the rejection rate increases to approximately 10--15\%, which is plausible given the greater estimation difficulty.

Additional results for $N=10$ and for the horseshoe prior are reported in Appendix~\ref{sec:ap:MI_sim_results}. We also provide results for differences between the overall posteriors obtained by pooling across the $m=100$ imputed datasets (i.e., after combining the corresponding posterior samples).

\subsection{Multiple Imputation -- Case Study}

To evaluate the proposed method on real-world data, we examine a travel time prediction problem. Our case study is inspired by the work of \cite{huang_travel_2020} and utilizes a dataset provided by the NYC Taxi and Limousine Commission (TLC) \citep{nyc_taxi_and_limousine_commission_tlc_tlc_2025}, containing yellow taxi trip records in New York City for the year 2016. The dataset includes information such as pickup and dropoff latitude and longitude, timestamps, number of passengers, and vendor ID. Additionally, it contains variables that are only known after trip completion, such as trip distance, trip duration, and payment amount.

We focus on a short-term travel time prediction task, in line with the forecasting approach of \cite{huang_travel_2020}. For this purpose, we use only variables available prior to the start of the trip: pickup and dropoff latitude and longitude. In addition, we employ the Open Source Routing Machine (OSM) \citep{dibbelt_customizable_2016}, which uses OpenStreetMap data for New York City to estimate trip duration and distance based on the trip’s start and end locations. These six variables serve as predictors in a Bayesian linear regression model, with the trip duration as the target variable. The model includes a log transformation of the target and scaling of all predictors, and we use the default priors provided by the \texttt{brms} \citep{burkner_brms_2024} package.

For our analysis, we select data from a single day --- Monday, March 7, 2016. The dataset is divided into 24 subsets, one for each hour of the day, based on the trip start time. Depending on the hour, the number of observations ranges from approximately 2,000 to 24,000. Each hourly subset contains some missing values, although the target variable (\textit{trip duration}) is always complete. The missing data in the predictors are partly due to failures in the OSM estimation process. The proportion of missing values per hourly dataset ranges from 1.1\% to 1.7\%.

\begin{figure*}[ht]
  \centering
  \subfigure[Number of log-probability (blue) and gradient log-probability evaluations (red) required to approximate the posteriors of $m=20$ imputed datasets. Error bars indicate the standard deviation across the 24 datasets representing the hours.]{\includegraphics[width=0.45\textwidth]{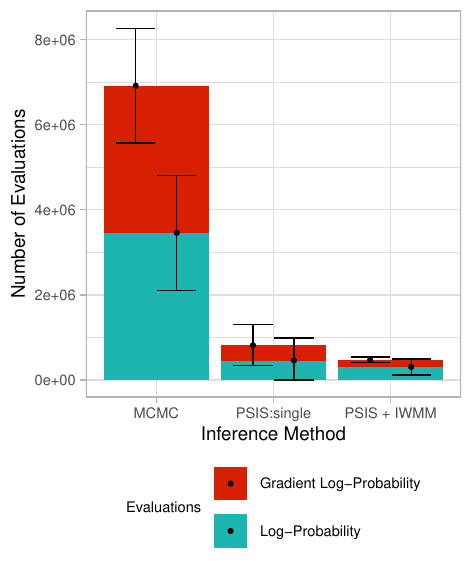}}
  \hspace{0.05\textwidth}
  \subfigure[Posterior distributions of model parameters to model the trip duration in the NYC taxi dataset, March 7, 2016, 12 pm, with $m=20$ imputed datasets for two different inference methods.]{\includegraphics[width=0.45\textwidth]{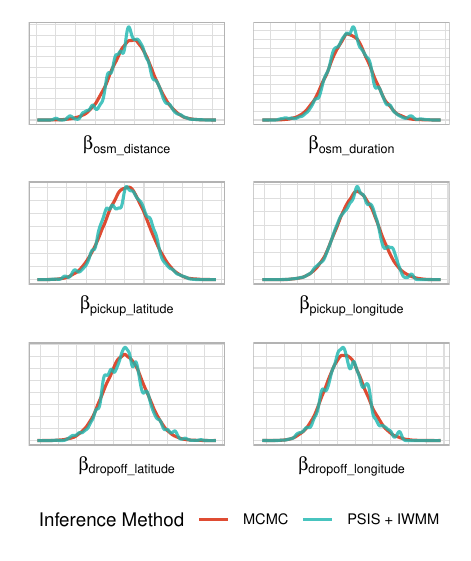}}
  \caption{Results of the case study with the real world NYC taxi dataset in the multiple imputation setting}
  \label{fig:NYC}
\end{figure*}

To account for uncertainty due to missing data, we perform multiple imputation using MICE \citep{buuren_mice_2011}, generating $m = 20$ imputed datasets for each hour. We use $m=20$ imputations, following a common rule of thumb \citep{white_multiple_2011}. Since fewer than 2\% of observations contain missing values, this choice appears reasonable for the present application, although no universal value of $m$ can be guaranteed to be sufficient in all settings. We then apply the iterative algorithm described in Section~\ref{sec:iterative_method} to approximate the marginalized posterior distributions of the model parameters, using both our full approach (\textit{PSIS+IWMM}) and a simplified variant without the IWMM step (\textit{PSIS:single}) that employs a single proposal distribution. As demonstrated in the simulation study in Section~\ref{sec:MI_sim_study}, incorporating a mixture proposal in the multiple imputation context did not offer significant advantages over using a single proposal. Therefore, we adopted the simpler approach in this analysis. For comparison, we also perform HMC sampling (brute-force \textit{MCMC}) using \texttt{brms} \citep{burkner_brms_2024, stan_development_team_stan_2024} on each of the 20 imputed datasets individually.

The results, in terms of log-probability and gradient log-probability evaluations, are shown in Figure~\ref{fig:NYC}(a). Notably, our proposed method (\textit{PSIS+IWMM}) requires only about 8\% of the log-probability evaluations and 5\% of the gradient evaluations compared to the baseline \textit{MCMC} approach. In contrast, the \textit{PSIS:single} variant requires, on average, 13\% of the log-probability and 10\% of the gradient evaluations relative to \textit{MCMC}. For each of the 24 hourly datasets, the complete iterative algorithm (\textit{PSIS+IWMM}) successfully approximated all 20 posterior distributions using resampling with PSIS and IWMM based on a single randomly selected proposal distribution.

Figure~\ref{fig:NYC}(b) presents the marginalized posterior distributions for one of these hourly datasets, specifically from 12–1 pm. The red line represents the density obtained from the brute-force \textit{MCMC} approach, marginalized over all $m=20$ imputed datasets, while the blue line shows the corresponding marginalized posterior density from our iterative method using PSIS and IWMM. The close alignment in shape and location of the distributions indicates that our method closely approximates the gold-standard HMC results. The bumpiness seen in the density estimate is mainly a consequence of marginalizing over a finite number of imputed datasets. Each imputed dataset induces a slightly different posterior distribution, and the marginalized posterior is obtained as a finite mixture of these imputation-specific posteriors. With $m=20$, this mixture can still display visible local irregularities, even when the main posterior features, such as location and spread, are well captured. Increasing the number of imputations would be expected to smooth this marginal approximation, but at the cost of additional runs of the imputation algorithm and corresponding second-step posterior computations.
Across all 24 hourly datasets and their $m=20$ imputations, the absolute difference in posterior means between the \textit{MCMC} baseline and \textit{PSIS+IWMM} are on the order of $10^{-4}$ when aggregated over all variables; the same holds for the posterior standard deviations. Using the MMD test \cite{gretton_kernel_2012}, only 1.25\% of the 480 evaluated datasets reject the null hypothesis that the \textit{MCMC} and \textit{PSIS+IWMM} samples are drawn from the same distribution.

\subsection{Surrogate Models -- Simulation Study \label{sec:surrogates_sim_study}}

To evaluate the performance of the iterative algorithm in a different context, we conduct a case study based on surrogate modeling. This study replicates and extends the work of \cite{reiser_uncertainty_2025}. There, a two-step procedure for uncertainty quantification in surrogate models was developed and tested in multiple experiments. One of these experiments is now re-examined here using the newly developed approach.

\subsubsection{Simulation Setup}
To demonstrate the performance of our method, we first consider a simplified example to avoid the complexities associated with real-life simulators and surrogates. Specifically, we examine a non-linear problem, using a one-dimensional logistic function as the simulator:
\begin{align}
    y = \mathcal{M}(\theta) = \frac{2}{1+\exp(-10\theta)} - 1 \,. \label{eq:true_simulator}
\end{align}
We apply two different surrogate models to approximate this simulator: a logistic surrogate, which includes the original simulator, and a PCE surrogate. 

\textbf{Logistic Surrogate}\quad For the logistic surrogate, we define
\begin{align}
    \widetilde{\mathcal{M}}_{\rm logistic}(\theta, \tau) = \frac{\tau_1}{1+\exp(-\tau_2(\theta - \tau_3))} + \tau_4 \,,
\end{align}
with surrogate parameters $\tau = [\tau_1, \tau_2, \tau_3, \tau_4]$. The original simulator is included in this surrogate model by fixing $\tau = [2, 10, 0, -1]$. 

For the training phase of the surrogate, we assign normal priors to the surrogate parameters with means equal to the true values $(2, 10, 0, -1)$ and the standard deviations set to $(1,10,1,1)$.

\textbf{PCE Surrogate}\quad To test a more challenging scenario, we also consider a surrogate model that does not include the true simulator. Specifically, we use a polynomial chaos expansion (PCE) model \citep{wiener_homogeneous_1938, oladyshkin_data-driven_2012, burkner_fully_2023}
\begin{align}
    \widetilde{\mathcal{M}}_{\rm PCE}(\theta, \tau) = \sum_{i=0}^d \tau_i \psi_i(\theta) \,,
\end{align}
where the parameters are ${\tau = [\tau_0, \tau_1, \dots, \tau_d]}$, and $\psi_i$ denote the Legendre polynomials \citep{sudret_global_2008}. For this study, we choose a polynomial degree of $d = 5$. Each coefficient $\tau_i$ is assigned a normal prior, $\tau_i \sim \mathcal{N}(0,5)$.

\textbf{Surrogate Training Step}\quad In the initial step, we generate $N_T = 10$ training data points using the simulator. The input values $\{\theta_T^{(n)}\}_{n=1}^{N_T}$ are selected as $N_T$ equally spaced points in the interval $[-1,1]$. The corresponding outputs $\{y_T^{(n)}\}_{n=1}^{N_T}$ are sampled from a normal distribution centered at the true simulator values with a standard deviation $\sigma = 0.01$, i.e., 
\begin{align}
    y_T^{(n)} \sim \mathcal{N}(\mathcal{M}(\theta_T^{(n)}), \sigma^2) \,.
\end{align}
    
To infer the surrogate parameters, we use MCMC implemented via the probabilistic programming language \texttt{Stan} \citep{stan_development_team_stan_2024}, using the \texttt{cmdstanr} package in \texttt{R} \citep{gabry_cmdstanr_2025}. Two chains are run for 1050 iterations each (including 1000 warm-up and 50 sampling iterations), resulting in $m = 100$ posterior draws ${\{\tau^{(i)}\}_{i=1}^m \sim p(\tau \mid D_T)}$. We use a relatively small number of posterior draws to enable a more direct comparison with the missing-data simulation study in Section~\ref{sec:MI_sim_study}. Since the choice of $m$ is difficult to determine in general, we also ran the same experiments with $m=1000$; the corresponding results are reported in Appendix~\ref{sec:ap:surrogate_sim_results} in Figure \ref{fig:surrogate_perc_MCMC_1000}.

\textbf{Inference Step}\quad For the inference phase, $N_I = 5$ measurement data points are generated from the true simulator
\begin{align}
    y_I \sim \mathcal{N}(\mathcal{M}(\theta_I^*), \sigma^2) \,,
\end{align}
where the true input is fixed at $\theta_I^* = -0.05$ and the observation noise is set to $\sigma = 0.01$. The prior for $\theta_I$ is a truncated normal distribution, ${\theta_I \sim \mathcal{N}_{[-1,1]}(0, 0.5^2)}$, while a uniform prior is chosen for the standard deviation, ${\sigma \sim \text{Uniform}(0, 0.05)}$.

\textbf{Simulation Parameters}\quad As in the multiple imputation setup, several methods are tested. The most straightforward approach involves fitting one separate model using HMC, an \textit{MCMC} method, for each of the $m = 100$ posterior draws $\{\tau^{(i)}\}_{i=1}^m$. In addition, we apply the iterative algorithm with PSIS only, using both a single proposal (\textit{PSIS:single}) and a mixture proposal constructed from five distributions (\textit{PSIS:mixture}), as well as the combined method of PSIS and IWMM (\textit{PSIS+IWMM}).

For selecting the representative draw used for the proposal distribution, multiple selection strategies are examined. These include selection a \textit{random} draw, choosing the draw with the highest estimated Pareto shape parameter $\hat{k}$ from the previous iteration, and the \textit{log-lik} method described in Section \ref{sec:special_case:surrogate}.
Additionally, an averaging method is tested, where the mean of the remaining posterior draws $ \{\tau^{(i)}\}_{i \in I} $ is used for the proposal. However, this approach resulted in an infinite loop in approximately 60\% of the tested settings and therefore is not considered in the subsequent results. The cause of this issue, arising when selecting a value $ \tau^* $ such that $ \tau^* \notin \{\tau^{(i)}\}_{i \in I} $, was discussed in Section \ref{sec:method_components}.

\textbf{Performance Evaluation}\quad 
To assess the performance of our method, we use the same metrics as those employed in the missing value simulation study presented in Section \ref{sec:MI_sim_study}.

\subsubsection{Results of the Simulation Study}

\begin{figure*}[t]
    \centering
    \includegraphics[width=\linewidth]{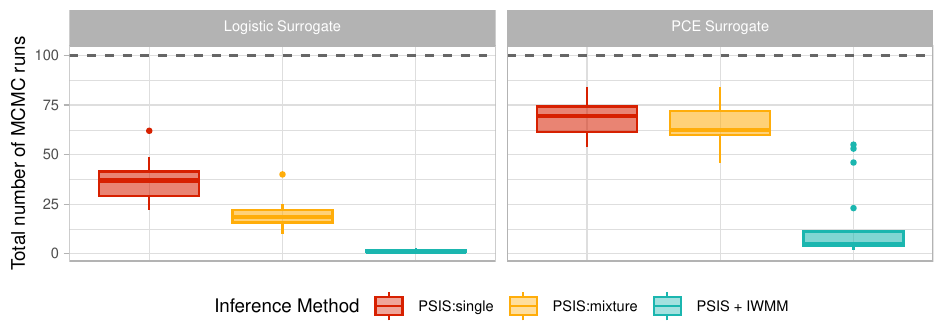}
    \caption{Number of MCMC runs required to approximate the posteriors of $ m = 100$ surrogate parameter draws, for the logistic and PCE surrogate (in the columns). Representative datasets were selected using the \textit{log-lik} procedure. The dashed horizontal line at 100 indicates the number of MCMC runs needed for the brute-force \textit{MCMC} approach.}
    \label{fig:no_MCMC_surrogate}
\end{figure*}

We first examine the number of MCMC runs required by the iterative algorithm, which are used to obtain the posterior distributions serving as proposal distributions in PSIS and IWMM. Since no substantial differences between selection strategies can be observed, presented results are from the \textit{log-lik} method only. Additional results for other selection methods can be found in Appendix \ref{sec:ap:surrogate_sim_results}. Figure~\ref{fig:no_MCMC_surrogate} displays the total number of MCMC runs, categorized by inference method and surrogate type (logistic and PCE). 
For the logistic surrogate, we observe that \textit{PSIS:single} requires the highest number of MCMC runs, with a median of 37. In contrast, for \textit{PSIS:mixture} only about 19 posterior distributions need to be fitted using MCMC. In most cases, the \textit{PSIS+IWMM} method finishes within 1 or 2 posteriors fitted with MCMC, effectively approximating all posterior distributions and thus requiring a significantly smaller number of MCMC runs.
The PCE surrogate represents a more challenging scenario because it does not include the true simulator and therefore inevitably introduces approximation error. This increases variability across the draws $\tau^{(i)}$ and, in turn, leads to greater heterogeneity in the corresponding posteriors $p(\theta\mid\tau^{(i)}, D_I)$, making approximation via PSIS and IWMM more difficult.
Both methods \textit{PSIS:single} and \textit{PSIS:mixture} perform similarly, with a median number of MCMC runs of 70 and 63 for the single and mixture proposals, respectively. Despite some outlier cases, the combined \textit{PSIS+IWMM} inference method consistently outperforms the other two in terms of minimizing the number of MCMC runs with a median of 5.

For $m=1000$ (see Figure~\ref{fig:surrogate_perc_MCMC_1000} in Appendix~\ref{sec:ap:surrogate_sim_results}), the results are even more favorable. On average, fewer than 6\% of the posteriors in the logistic-surrogate setting need to be computed directly via MCMC, and fewer than 20\% in the PCE-surrogate setting.

We also assessed how closely the resampled posterior distributions obtained with \textit{PSIS+IWMM} match the gold-standard \textit{MCMC} baseline. Table~\ref{tab:surrogate_posterior_distances} reports the absolute differences in posterior means and standard deviations, averaged over 20 datasets. For the logistic surrogate, agreement is excellent: posterior means and standard deviations are nearly identical. In contrast, discrepancies are larger for the PCE surrogate. This is likely because the PCE surrogate has more parameters and therefore wider first-step posteriors, which propagates into greater variability in the resulting second-step posterior approximations. 
When comparing posterior means obtained from MCMC for the individual posteriors $p(\theta_I \mid \tau^{(i)}, D_I)$ across different realizations $\tau^{(i)}$, the logistic-surrogate means differ by an average absolute pairwise distance of about 0.0028 (for $m=100$). For the PCE surrogate, the corresponding average absolute difference is 0.0664. Hence, the PCE posteriors are substantially more dispersed (i.e., less similar across $\tau^{(i)}$), which helps explain why approximating them with PSIS or IWMM is more challenging.

\begin{table}[ht]
    \centering
    \caption{Distances between the posterior distributions $p(\theta_I \mid D_T, D_I)$ of $\theta_I$ obtained via \textit{MCMC} and via the iterative algorithm \textit{PSIS+IWMM}. Values are the absolute differences in posterior mean and standard deviation, averaged over 20 datasets.}
    \begin{tabular}{cccc}
    \hline
       $m$ & Surrogate   &  Mean  & Standard Deviation\\ \hline
      \multirow{2}{*}{$100$} &  Logistic  & 8.0471e-05 & 0.0001 \\
        & PCE  &  0.0178 & 0.0528 \\ \hline
      \multirow{2}{*}{$1000$} &  Logistic  &  0.0004 & 0.0005 \\
        & PCE  &  0.0698  &  0.0834\\ \hline
    \end{tabular}
    \label{tab:surrogate_posterior_distances}
\end{table}

\begin{figure*}[ht]
    \centering
    \includegraphics[width=\linewidth]{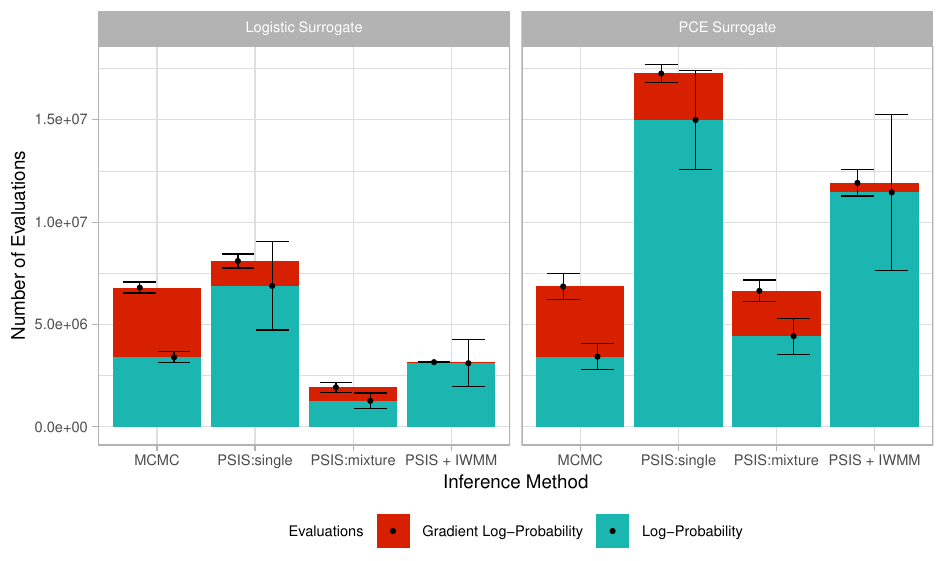}
    \caption{Number of log-probability (blue) and gradient log-probability evaluations (red) required to approximate the posteriors of $m = 100$ surrogate parameter draws, for the logistic and the PCE surrogate (in the columns). Representative draws were selected using the \textit{log-lik} method. Error bars indicate the standard deviation across repeated runs of the same setup.}
    \label{fig:barplot_evaluation_surrogate}
\end{figure*}

To further evaluate performance, we analyze the number of log-probability and gradient evaluations required by each inference method, relative to running brute-force \textit{MCMC}. Figure~\ref{fig:barplot_evaluation_surrogate} displays stacked bar plots for the different inference methods, with results for the logistic surrogate shown on the left and the PCE surrogate on the right. To better illustrate the relative efficiency of the proposed methods compared to the baseline \textit{MCMC} approach, Table~\ref{tab:surrogate_performance} reports the corresponding ratios of required evaluations.

\begin{table*}
    \centering
    \resizebox{\textwidth}{!}{
    \begin{tabular}{c|c c c | c c c}
         & \multicolumn{3}{c|}{Logistic Surrogate} & \multicolumn{3}{c}{PCE Surrogate} \\
         & PSIS:single & PSIS:mixture & PSIS+IWMM &  PSIS:single & PSIS:mixture & PSIS+IWMM\\ \hline
        log-probability & 2.02 & 0.37 & 0.91  &    4.37 & 1.29 & 3.34 \\
        gradient log-prob. & 0.35 & 0.20 & 0.01 &     0.66 & 0.65 & 0.14 \\
    \end{tabular}
    }
    \vspace{0.8em}
    \caption{Ratio of the mean number of log-probability and gradient log-probability evaluations for each inference method (columns), relative to the brute-force \textit{MCMC} approach that runs HMC separately on all $m = 100$ surrogate parameter draws. A ratio of 1 indicates equivalent computational effort; values below 1 indicate fewer evaluations and thus greater computational efficiency.}
    \label{tab:surrogate_performance}
\end{table*}

For the logistic surrogate, the \textit{PSIS:single} requires slightly more log-probability evaluations than the baseline \textit{MCMC}, but significantly fewer gradient evaluations, which are typically more computationally expensive. The \textit{PSIS:mixture} inference method, as well as \textit{PSIS+IWMM}, succeed in approximating all posterior distributions with fewer evaluations of both log-probability and gradient evaluations.
In contrast, the PCE surrogate poses a greater challenge, as the posterior distributions $p(\theta_I \mid \tau^{(i)}, D_I)$ differ more substantially. In the worst case --- observable in \textit{PSIS:single} --- the iterative algorithm requires more log-probability evaluations than baseline \textit{MCMC}. This occurs because the algorithm needs many iterations to evaluate all posterior distributions. During each iteration, importance ratios are computed for each of the remaining posteriors. If PSIS fails in many of these cases, a significant number of importance ratio calculations are only used to determine that PSIS does not work for this particular proposal-target combination, rather than approximating the target distribution. As a result, the number of log-probability evaluations exceeds the ones of running \textit{MCMC} separately. In contrast, the number of gradient evaluations required by the new method cannot exceed those needed for running brute-force \textit{MCMC}, as they are only necessary for calculating the proposal in the iterative algorithm with HMC, and there can be at most $m=100$ different proposals.
The \textit{PSIS:mixture} method substantially improves PSIS performance compared to the single proposal with only having a few more log-probability evaluations than the \textit{MCMC} method.
The combined \textit{PSIS+IWMM} approach again results in the fewest gradient evaluations, indicating that only a small number of posteriors needed to be computed using HMC. The high number of log-probability evaluations in this method arises from applying the IWMM algorithm (see Section \ref{sec:method_components}) in each iteration of the iterative algorithm for multiple different target distributions.

In conclusion, the proposed method improves the efficiency of posterior approximation for computing the marginalized posterior via the Monte Carlo estimate. In the simpler setting --- using the logistic surrogate, which incorporates the true simulator --- the method can approximate all posterior distributions with only one or two expensive HMC runs. In the more complex scenario involving the PCE surrogate, the method faces greater challenges, but still achieves notable reductions in computational effort. The number of gradient log-probability evaluations can be reduced to less than 14\% of those required by the brute-force \textit{MCMC} approach, as shown in Table~\ref{tab:surrogate_performance}. Since gradient evaluations typically dominate the overall computational cost, our proposed method achieves improved efficiency even in more complex settings.

\section{Conclusion}

In this work, we introduced a diagnostic-driven and computationally efficient method for uncertainty propagation in Bayesian two-step modeling frameworks. Our approach combines Pareto Smoothed Importance Sampling (PSIS) and Importance Weighted Moment Matching (IWMM) to substantially reduce the number of costly sampling runs typically required for reliable uncertainty propagation. Rather than performing separate MCMC simulations for each of the multiple realizations of the uncertainty quantity, we run MCMC for only a small, carefully selected subset of representative draws and approximate the remaining posteriors via the two importance sampling techniques. We focus on two important application domains where uncertainty propagation is particularly critical: multiple imputation and surrogate modeling.

Through simulation studies in these two application settings, as well as a real-world case study, we found that the method closely approximates the HMC-based posterior summaries in many of the settings considered while yielding substantial computational savings. Specifically, it reduces the number of log-probability evaluations by up to 90\% and the number of gradient evaluations by 85–99\% compared to a brute-force HMC baseline. The accuracy of the resulting marginalized posteriors is maintained through the use of diagnostics that validate each approximation, providing a mechanism for identifying cases in which close agreement with the HMC reference is expected, and for flagging cases in which the approximation may be unreliable.

To support efficient uncertainty propagation, we also developed and evaluated several strategies for selecting representative values used to construct proposal distributions for PSIS and IWMM. In our simulation studies, the choice of selection strategy had only a limited impact on the performance metrics considered. Notably, even a simple random selection performed comparably to more sophisticated approaches. This suggests that either the choice of representative values is not critical to the method’s effectiveness, or that the current selection techniques are not sufficiently sensitive to differences in the posterior space structure. Consequently, extensive computational effort for optimizing representative value selection may not be warranted in settings similar to those studied here.
Despite its good performance in many of the scenarios considered, the method showed limitations in more challenging settings. For example, in missing data problems characterized by complex posterior structure --- such as those arising from shrinkage priors, high-dimensional parameter spaces, and a high proportion of missingness --- the affine transformations used in IWMM are insufficient. In such cases, the discrepancy between the target and proposal distribution leads to degraded approximation quality. More expressive transformation techniques may be required, similar to those needed in other applications of IWMM, such as leave-one-group-out cross-validation \citep{roberts_cross-validation_2017}.

Although we demonstrated the method in two specific domains, the same computational idea may be applicable to other modeling scenarios with a two-step structure in which uncertainty propagation is essential. Bayesian cut models are one example of such a setting. Developing scalable and user-friendly implementations will be crucial for broader adoption. While the current implementation is mostly written in \texttt{R} (see \href{https://github.com/sjedhoff/efficient-2step-uncertainty-paper}{GitHub}), versions in a compiled language such as \texttt{C++}, or full integration with probabilistic programming frameworks like \texttt{Stan} could greatly enhance practical performance.

Choosing $m$, the number of first-step draws $\tau^{(i)}$, is inherently context-dependent and often difficult in practice. Even in the closely related importance-sampling literature, the choice of $m$ has been debated for decades, underscoring that there is no universally optimal rule. Our approach alleviates this trade-off by making it feasible to use larger values of $m$ and thus propagate first-step uncertainty more faithfully, without incurring a proportional increase in computational cost.

Overall, this work presents a flexible and computationally efficient framework for uncertainty propagation in Bayesian two-step procedures, providing a practical route to improved scalability while retaining good agreement with HMC-based reference results in the examples studied.

\backmatter

\bmhead{Acknowledgements}

This research was supported by the German Research Foundation (DFG) via Collaborative Research Center 391 "Spatio-Temporal Statistics for the Transition of Energy and Transport" -- 520388526.

\onecolumn
\begin{appendices}

\FloatBarrier
\section{Notation}

\begin{table}[ht]
    \centering
    \begin{tabular}{ll}
        \textbf{Symbol} & \textbf{Description} \\ \toprule
        $\theta$  & variable of interest $\theta \sim p(\theta \mid C)$ \\
        $\tau$ & underlying variable with $\tau \sim p(\tau \mid C)$ \\
        $\tau^{(i)}$ & draws of $p(\tau)$ in the first step, $i=1, \dots, m$ \\
        $p(\theta \mid \tau^{(i)})$ & posterior distribution corresponding to draw $\tau^{(i)}$ -- target distribution \\
        $\tau^*$ & selected draw from  $\tau^{(1)}, \dots, \tau^{(m)}$\\
        $p(\theta \mid \tau^*)$ & posterior distribution corresponding to selected draw $\tau^*$ -- proposal distribution \\
        $\theta^{(s)}_*$ & draws from the proposal distribution $p(\theta \mid \tau^*)$, $s=1, \dots, S$ \\
        $w_i(\theta_*^{(s)})$ & importance weight with target $p(\theta\mid\tau^{(i)})$ and proposal $p(\theta\mid\tau^*)$ evaluated at draw $\theta_*^{(s)}$ \\
        $\hat{k}_i$ & estimated Pareto shape parameter for PSIS with target $p(\theta\mid\tau^{(i)})$ -- used as a diagnostic metric\\
        $\breve{\theta}^{(s)}_*$ & transformed draws through IWMM from the proposal distribution $p(\theta \mid \tau^*)$, $s=1, \dots, S$ \\
        $\hat{k}_{\text{MM}, i}$ &  estimated Pareto shape parameter for IWMM with target $p(\theta \mid \tau^{(i)})$ \\
        $p_{\rm mix}(\theta \mid \{\tau^{(j)}\})$ & mixture distribution used as the proposal \\
        \hline
       \multicolumn{2}{l}{Missing Data Problem}  \\ \hline 
        $D$ & dataset containing N observations of target $y$ and $p$ covariates $X = (x_1, \dots, x_p)$ \\
        $D_{\rm obs} = (y_{\rm obs}, X_{\rm obs})$ & observed dataset \\
        $\theta$ & regression model parameters; describing relationsship between $y$ and $X$ \\
        $D^{(i)}$ & imputed dataset, $i=1, \dots, m$ \\
        $p_{\rm MICE}(\theta \mid D_{\rm obs})$ & density of interest \\
        $J^*$ & index set that indicates in which observations the original dataset $D$ had missing values \\
        \hline
        \multicolumn{2}{l}{Surrogate Modeling} \\ \hline
        $\mathcal{M}$ & simulator with input $\theta$ \\
        $\widetilde{\mathcal{M}}$ & surrogate model with input $\theta$ and trainable surrogate parameters $\tau$ \\
        $\theta$ & simulator input parameter \\
        $D_T = \{\theta_T^{(n)}, y_T^{(n)} \}_{n=1}^{N_T}$ & training data produced by the simulator \\
        $\{ \tau^{(i)}\}_{i=1}^m$ & $m$ samples of the posterior distribution $p(\tau | D_T)$ \\
        $\theta_I$ & true simulator input variable \\
        $D_I = \{y_I^{(n)} \}_{n=1}^{N_I}$ & real world measurements produced by the simulator with input $\theta_I$ \\
        
        \bottomrule
    \end{tabular}
    \label{tab:notation}
\end{table}

\FloatBarrier
\onecolumn
\section{Iterative Algorithm for using a mixture proposal \label{sec:ap:algo_mix}}
\begin{algorithm}[ht]
\caption{Iterative Algorithm for Efficiently Calculating the Marginalized Posterior (Mixture Proposal)}\label{algo:method_mix}
\begin{algorithmic}[1]
\Require Variable of interest $\theta \sim p(\theta)$, underlying variable $\tau \sim p(\tau)$, mixture size $I_{\text{mix}}$
\State Draw $m$ samples $\tau^{(1)}, \dots, \tau^{(m)} \sim p(\tau)$
\State Set index set $I = \{1,\dots, m\}$
\While{$|I| > |I_{\text{mix}}|$}
    \State Select $|I_{\text{mix}}|$ representative values from $\{\tau^{(i)}\}_{i \in I}$ 
    \State Run MCMC to access their posterior distributions: $\{\theta^{(s)}_i\}_{s=1}^S$
    \State Build the mixture: 
    \[
    p_{\text{mix}}(\theta \mid \{\tau^{(i)}\}) = \frac{1}{|I_{\text{mix}}|} \sum_{i \in I_{\text{mix}}} p(\theta \mid \tau^{(i)})
    \]
    \State Obtain samples $\{\theta^{(s)}_*\}_{s=1}^S \sim p_{\text{mix}}(\theta \mid \{\tau^{(i)}\})$ via resampling
    \State Remove the $|I_{\text{mix}}|$ selected values from $I$
    \For{$i \in I$}
        \State Run PSIS with target $p(\theta \mid \tau^{(i)})$ and proposal $p_{\text{mix}}(\theta \mid \{\tau^{(i)}\})$
        \State Obtain importance weights and diagnostic metric $\hat{k}_i$
        \If{$\hat{k}_i < 0.7$}
            \State Use the importance weights and $\{\theta^{(s)}_*\}$ to resample posterior draws for $p(\theta \mid \tau^{(i)})$
            \State Remove $i$ from $I$: $I \gets I \setminus \{i\}$
        \EndIf
    \EndFor
\EndWhile
\For{$i \in I$}
    \State Run MCMC to obtain posterior draws from $p(\theta \mid \tau^{(i)})$
\EndFor
\State Compute marginalized posterior via Monte Carlo estimate: $p(\theta \mid C) \approx \frac{1}{m} \sum_{i=1}^m p(\theta \mid \tau^{(i)})$
\end{algorithmic}
\end{algorithm}

\FloatBarrier
\onecolumn
\section{Neural variational inference for cut feedback in the missing data setting \label{sec:ap:nevicut}}
\setcounter{figure}{11}   
\setcounter{table}{2}
\renewcommand{\thefigure}{\arabic{figure}}
\renewcommand{\thetable}{\arabic{table}}

The missing data example considered in this paper can also be interpreted from the perspective of modular Bayesian inference and cut feedback. In this formulation, the imputation model constitutes an upstream module, while the substantive regression model constitutes a downstream module.
Let $D_\mathrm{obs}$ denote the observed incomplete dataset and let $D_\mathrm{mis}$ denote the missing entries in the dataset (see Section \ref{sec:special_case:MI} for more details of the setup). The imputation procedure provides draws
\begin{align*}
    D_{\mathrm{mis}}^{(m)} \sim p_{\mathrm{imp}}(D_{\mathrm{mis}} \mid D_{\mathrm{obs}}),
\qquad m = 1,\ldots,M,
\end{align*}
which induce complete datasets $D^{(m)} = (D_{\mathrm{obs}}, D_{\mathrm{mis}}^{(m)})$.
In the notation of cut models, the missing values $D_\mathrm{mis}$ play the role of the upstream quantity, while the regression parameters $\theta$ are the downstream parameters of interest.

The corresponding cut posterior can be written as
\begin{align*}
    p_{\mathrm{cut}}(D_{\mathrm{mis}}, \theta \mid D_{\mathrm{obs}})
 = p_{\mathrm{imp}}(D_{\mathrm{mis}} \mid D_{\mathrm{obs}})
p(\theta \mid D_{\mathrm{obs}}, D_{\mathrm{mis}}).
\end{align*}
The first factor describes uncertainty about the missing values as represented by the imputation model. The second factor is the posterior distribution of the regression parameters conditional on one completed dataset. The cut indicates that inference in the downstream regression model is not allowed to update the imputation distribution. Thus, uncertainty is propagated from the imputation model to the regression model, but feedback from the regression model to the imputation model is prevented.

This interpretation connects the missing data problem to the neural variational cut-Bayes approach of \citet{song_neural_2025}. In their framework, samples from the upstream distribution are used directly, and a conditional variational approximation is learned for the downstream posterior. Applied to the present missing data setting, the upstream samples are the imputed missing values $\eta^{(m)} = D_{\mathrm{mis}}^{(m)},$ and the downstream target is $p(\theta \mid D_{\mathrm{obs}}, \eta^{(m)})$.
Instead of fitting the downstream model separately for every imputed dataset, one can train a conditional variational approximation
\begin{align*}
    q_{\lambda}(\theta \mid \eta)
\approx
p(\theta \mid D_{\mathrm{obs}}, \eta),
\end{align*}
where $q_{\lambda}$ is parameterized by a flexible neural density estimator, such as a normalizing flow.

The NeVI-Cut \citep[Neural Variational Inference for Cut-Bayes,][]{song_neural_2025} formulation is applicable in this setting because the multiple imputation procedure already provides samples from an upstream uncertainty distribution, and the downstream regression model can be evaluated conditionally on each completed dataset. However, its use also changes the nature of the approximation. The quality of the result depends on the expressiveness of the variational family and on the success of the optimization procedure. Moreover, for a fair comparison with MCMC-based approaches, the same downstream likelihood and priors should be used.

To compare the performance of the NeVI-Cut method by \cite{song_neural_2025} to our proposed method, we use the Gaussian regression model from the simulation study in Section~\ref{sec:MI_sim_study}.
The downstream parameter vector is $\theta = (\beta_0,\beta_1,\ldots,\beta_p,\log \sigma)$, and the likelihood conditional on a completed dataset is $y_i \mid x_i, \theta \sim\mathcal{N}\left(\beta_0 + x_i^\top \beta, \sigma^2\right), i=1,\ldots,n.$
The variational approximation is trained using the imputed values $\{ \eta^{(m)} \}_{m=1}^M$ as conditional variables. After training, conditional posterior draws can be generated for each imputation,
\begin{align*}
    \theta^{(s,m)} \sim q_{\lambda}(\theta \mid \eta^{(m)}),
\qquad s=1,\ldots,S,\quad m=1,\ldots,M.
\end{align*}
These draws provide a variational approximation to the same collection of imputation-specific posterior distributions that are approximated by the iterative method introduced in Section~\ref{sec:iterative_method}. To ensure comparability, we use the same imputed datasets for both approaches and compare the posterior samples obtained from NeVI-Cut with the MCMC baseline using the same evaluation criteria as for the iterative method.

\begin{figure}
    \centering
    \includegraphics[width=\linewidth]{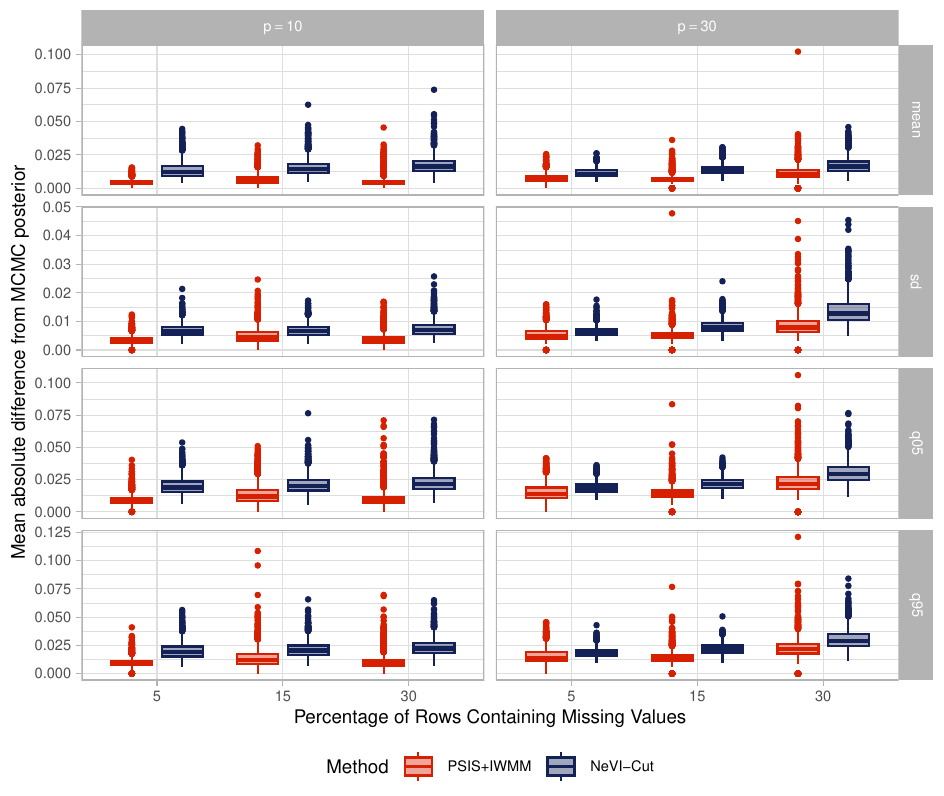}
    \caption{Mean absolute differences between posterior summaries obtained from \textit{MCMC} and from the \textit{PSIS+IWMM} and \textit{NeVI-Cut} resampled posterior for datasets of size $N=100$. For each missingness level (5\%, 15\%, and 30\% of rows containing missing values), boxplots summarize the distribution across the 20 datasets and their 100 imputed datasets; within each dataset, differences are averaged over the $p$ parameters. Rows indicate the summary metric (posterior mean, standard deviation, 5\% quantile, and 95\% quantile). Results are shown for $p=10$ (left) and $p=30$ (right).}
    \label{fig:nevicut_mcmc_diff}
\end{figure}

As in Section~\ref{sec:MI_sim_study}, we compare posterior means, standard deviations, and the 5\% and 95\% quantiles in terms of mean absolute differences from the corresponding MCMC posterior summaries. Figure~\ref{fig:nevicut_mcmc_diff} shows the results for datasets of size $N=100$ under the standard prior specification. Across all four posterior summary statistics, the mean absolute differences from the MCMC baseline are, on average, smaller for the iterative method based on PSIS+IWMM than for NeVI-Cut. This can be seen from the lower positions of the boxes and medians in the corresponding boxplots.

A comparison based on the number of log-probability and gradient evaluations is less straightforward for NeVI-Cut than for MCMC or PSIS+IWMM. NeVI-Cut is trained by solving an optimization problem in which the variational objective is evaluated repeatedly. During backpropagation, gradients are computed with respect to the neural-network or normalizing-flow parameters $\lambda$, rather than with respect to the downstream model parameters $\theta$. These gradients therefore correspond to a different computational object than the gradients used in HMC. For this reason, gradient-evaluation counts for NeVI-Cut are not directly comparable to the gradient-evaluation counts reported for HMC-based posterior simulation.

This experiment illustrates that methods developed for cut-feedback problems can, in principle, be applied to the two-step uncertainty propagation problems considered in this paper. In the missing-data setting studied here, however, the proposed iterative method provides posterior samples that are closer to the MCMC baseline than those obtained from the NeVI-Cut implementation of \citet{song_neural_2025}, according to the posterior-summary discrepancies considered above.

\FloatBarrier
\onecolumn
\section{Additional Results of the Simulation Study for the Multiple Imputation Problem \label{sec:ap:MI_sim_results}}

\begin{figure}
    \centering
    \includegraphics[width=\linewidth]{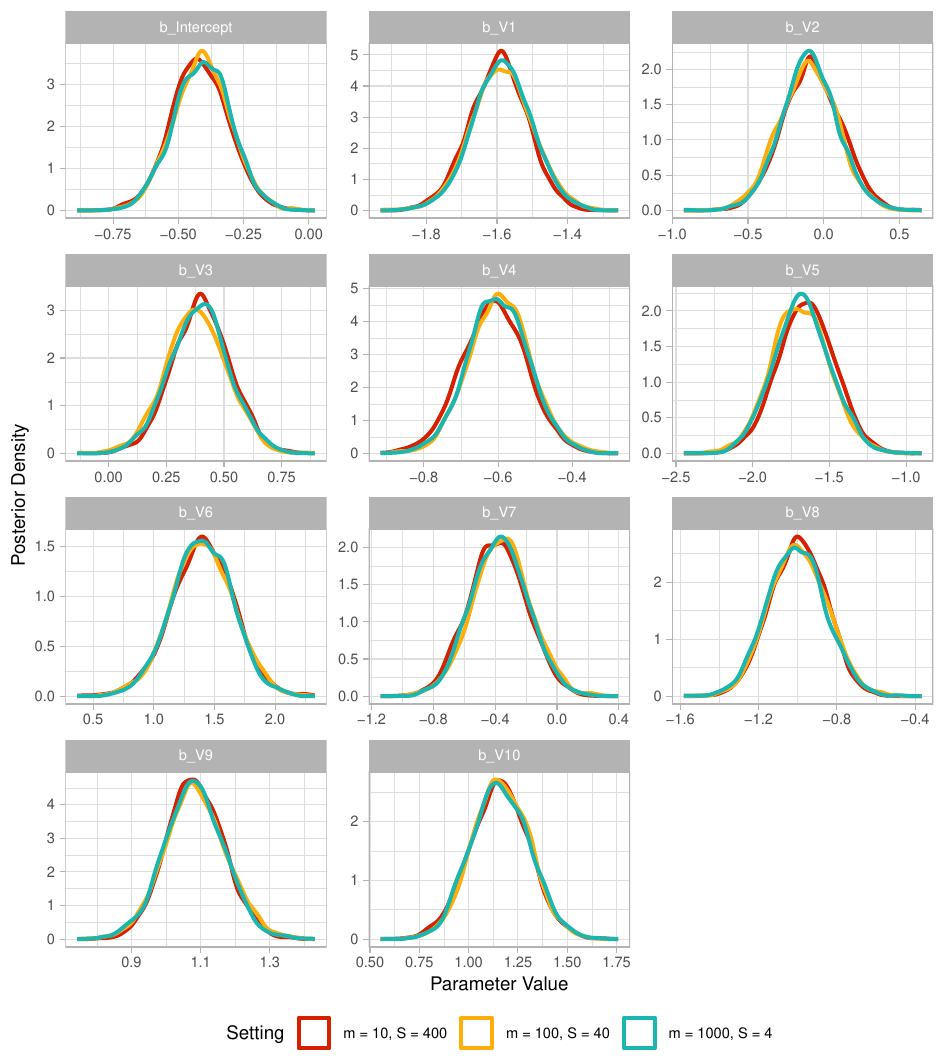}
\caption{Posterior density estimates (MCMC) for the regression coefficients ($N=100,p=10$) under three allocations of a fixed total posterior sample size (\(m\cdot S=4000\)) across multiple imputations: $m=10,\,S=400$, $m=100,\,S=40$, and $m=1000,\,S=4$. Each panel corresponds to one parameter ($b_{\text{Intercept}}, b_{V1}, \ldots, b_{V10}$). The near-complete overlap of the curves indicates that, in this example, the inferred posterior is largely insensitive to the specific choice of $m$ and $S$ when the total number of draws is held constant.}
    \label{fig:posteriors_vary_mS}
\end{figure}

\begin{table}[ht]
\centering
\caption{Differences between the individual posterior distributions obtained via MCMC and via the iterative algorithm under the standard prior. Values are the mean absolute differences in the corresponding metric, averaged over 20 datasets, with $m = 100$ imputations per dataset. For each dataset, the two sets of posterior samples are compared using the kernel maximum mean discrepancy (MMD) test at significance level $\alpha = 0.05$. The far-right column reports the percentage of datasets in which the null hypothesis (that the two sample sets are drawn from the same distribution) is rejected.}
\label{tab:mi_difference_posterior_standard}
\begin{tabular}{rrrrrrrr}
  \hline
$N$ & $p$ & $\%$ missing rows & Mean & Standard Deviation & $5\%$ Quantile & $95\%$ Quantile & \% MMD rejection \\ 
  \hline
\multirow{6}{*}{10} & \multirow{3}{*}{2} & 5 & 0.0188 & 0.0177 & 0.0451 & 0.0427 & 0.45 \\
   &  & 15 & 0.0241 & 0.0239 & 0.0582 & 0.0578 & 4.35 \\
   &  & 30 & 0.0227 & 0.0218 & 0.0532 & 0.0517 & 3.10 \\\cmidrule{2-8} 
   & \multirow{3}{*}{5} & 5 & 0.0432 & 0.0669 & 0.1297 & 0.1213 & 11.90 \\ 
   &  & 15 & 0.0671 & 0.0828 & 0.1520 & 0.1622 & 15.30 \\ 
   &  & 30 & 0.0624 & 0.0719 & 0.1414 & 0.1465 & 14.65 \\ \hline
  \multirow{6}{*}{100} & \multirow{3}{*}{10} & 5 & 0.0045 & 0.0036 & 0.0095 & 0.0097 & 0.30 \\ 
   &  & 15 & 0.0069 & 0.0051 & 0.0137 & 0.0141 & 6.00 \\ 
   &  & 30 & 0.0051 & 0.0038 & 0.0106 & 0.0104 & 4.45 \\ \cmidrule{2-8} 
   & \multirow{3}{*}{30} & 5 & 0.0077 & 0.0054 & 0.0155 & 0.0157 & 3.75 \\ 
   &  & 15 & 0.0069 & 0.0053 & 0.0145 & 0.0146 & 1.90 \\ 
   &  & 30 & 0.0114 & 0.0000 & 0.0229 & 0.0227 & 10.45 \\ 
   \hline
\end{tabular}
\end{table}

\begin{table}[ht]
\centering
\caption{Differences between the individual posterior distributions obtained via MCMC and via the iterative algorithm under the horseshoe prior. Values are the mean absolute differences in the corresponding metric, averaged over 20 datasets, with $m = 100$ imputations per dataset. For each dataset, the two sets of posterior samples are compared using the kernel maximum mean discrepancy (MMD) test at significance level $\alpha = 0.05$. The far-right column reports the percentage of datasets in which the null hypothesis (that the two sample sets are drawn from the same distribution) is rejected.}
\label{tab:mi_difference_posterior_horseshoe}
\begin{tabular}{rrrrrrrr}
  \hline
$N$ & $p$ & $\%$ missing rows & Mean & Standard Deviation & $5\%$ Quantile & $95\%$ Quantile & \% MMD rejection \\ 
  \hline
\multirow{6}{*}{10} & \multirow{3}{*}{2} & 5 & 0.0156 & 0.0140 & 0.0365 & 0.0331 & 0.45 \\ 
   &  & 15 & 0.0312 & 0.0181 & 0.0482 & 0.0527 & 5.20 \\ 
   &  & 30 &0.0249 & 0.0167 & 0.0454 & 0.0464 & 6.70 \\ \cmidrule{2-8} 
   & \multirow{3}{*}{5} & 5 & 0.0355 & 0.0249 & 0.0651 & 0.0554 & 13.35 \\ 
   &  & 15 &0.0493 & 0.0333 & 0.0818 & 0.0875 & 23.45 \\ 
   &  & 30 & 0.0575 & 0.0365 & 0.0946 & 0.0900 & 27.90 \\  \hline
  \multirow{6}{*}{100} & \multirow{3}{*}{10} & 5 & 0.0044 & 0.0033 & 0.0094 & 0.0095 & 0.65 \\ 
   &  & 15 & 0.0086 & 0.0059 & 0.0171 & 0.0164 & 12.10 \\ 
   &  & 30 & 0.0107 & 0.0066 & 0.0186 & 0.0190 & 38.40 \\ \cmidrule{2-8} 
   & \multirow{3}{*}{30} & 5 &  0.0088 & 0.0060 & 0.0175 & 0.0176 & 8.45 \\ 
   &  & 15 &0.0086 & 0.0048 & 0.0147 & 0.0149 & 36.25 \\ 
   &  & 30 &  0.0001 & 0.0000 & 0.0001 & 0.0001 & 0.15 \\ 
   \hline
\end{tabular}
\end{table}

\begin{table}[ht]
\centering
\caption{Differences between the final posterior distributions obtained via MCMC and via the iterative algorithm under the standard prior. Values are the mean absolute differences in the corresponding metric, averaged over 20 datasets. For each dataset, the two sets of posterior samples are compared using the kernel maximum mean discrepancy (MMD) test at significance level $\alpha = 0.05$. The far-right column reports the percentage of datasets in which the null hypothesis (that the two sample sets are drawn from the same distribution) is rejected.}
\label{tab:mi_difference_posterior_standard_total}
\begin{tabular}{rrrrrrrr}
  \hline
$N$ & $p$ & $\%$ missing rows & Mean & Standard Deviation & $5\%$ Quantile & $95\%$ Quantile & \% MMD rejection \\ 
  \hline
\multirow{6}{*}{10} & \multirow{3}{*}{2} & 5 & 0.0141 & 0.0127 & 0.0357 & 0.0299 & 0 \\ 
   &  & 15 &0.0155 & 0.0166 & 0.0370 & 0.0394 & 0 \\ 
   &  & 30 & 0.0127 & 0.0158 & 0.0309 & 0.0335 & 0 \\  \cmidrule{2-8} 
   & \multirow{3}{*}{5} & 5 & 0.0329 & 0.0554 & 0.1033 & 0.0879 & 0 \\ 
   &  & 15 &0.0409 & 0.0595 & 0.0868 & 0.1082 & 0 \\ 
   &  & 30 & 0.0357 & 0.0440 & 0.0763 & 0.0792 & 0 \\  \hline
  \multirow{6}{*}{100} & \multirow{3}{*}{10} & 5 &0.0029 & 0.0025 & 0.0061 & 0.0066 & 0 \\ 
   &  & 15 &0.0034 & 0.0031 & 0.0070 & 0.0076 & 0 \\ 
   &  & 30 &  0.0027 & 0.0021 & 0.0049 & 0.0048 & 0 \\   \cmidrule{2-8} 
   & \multirow{3}{*}{30} & 5 &  0.0042 & 0.0032 & 0.0083 & 0.0083 & 0 \\ 
   &  & 15 &0.0034 & 0.0030 & 0.0066 & 0.0067 & 0 \\ 
   &  & 30 & 0.0043 & 0.0045 & 0.0088 & 0.0088 & 0 \\ 
   \hline
\end{tabular}
\end{table}

\begin{table}[ht]
\centering
\caption{Differences between the final posterior distributions obtained via MCMC and via the iterative algorithm under the horseshoe prior. Values are the mean absolute differences in the corresponding metric, averaged over 20 datasets. For each dataset, the two sets of posterior samples are compared using the kernel maximum mean discrepancy (MMD) test at significance level ($\alpha = 0.05$). The far-right column reports the percentage of datasets in which the null hypothesis (that the two sample sets are drawn from the same distribution) is rejected.}
\label{tab:mi_difference_posterior_horseshoe_total}
\begin{tabular}{rrrrrrrr}
  \hline
$N$ & $p$ & $\%$ missing rows & Mean & Standard Deviation & $5\%$ Quantile & $95\%$ Quantile & \% MMD rejection \\ 
  \hline
\multirow{6}{*}{10} & \multirow{3}{*}{2} & 5 & 0.0109 & 0.0100 & 0.0273 & 0.0223 & 0 \\ 
   &  & 15 &0.0240 & 0.0119 & 0.0317 & 0.0360 & 0 \\ 
   &  & 30 &0.0183 & 0.0095 & 0.0328 & 0.0274 & 5 \\ \cmidrule{2-8} 
   & \multirow{3}{*}{5} & 5 &0.0281 & 0.0169 & 0.0466 & 0.0348 & 5 \\  
   &  & 15 &0.0339 & 0.0210 & 0.0490 & 0.0529 & 0 \\ 
   &  & 30 & 0.0386 & 0.0256 & 0.0557 & 0.0592 & 10 \\  \hline
  \multirow{6}{*}{100} & \multirow{3}{*}{10} & 5 &0.0028 & 0.0023 & 0.0062 & 0.0062 & 0 \\ 
   &  & 15 & 0.0039 & 0.0030 & 0.0082 & 0.0078 & 0 \\
   &  & 30 &  0.0026 & 0.0023 & 0.0052 & 0.0047 & 0 \\  \cmidrule{2-8} 
   & \multirow{3}{*}{30} & 5 &  0.0042 & 0.0030 & 0.0085 & 0.0085 & 0 \\ 
   &  & 15 &0.0017 & 0.0013 & 0.0033 & 0.0031 & 0 \\ 
   &  & 30 &   0.0001 & 0.0000 & 0.0001 & 0.0001 & 0 \\ 
   \hline
\end{tabular}
\end{table}

\begin{figure*}[ht]
    \centering
    \includegraphics[width=\linewidth]{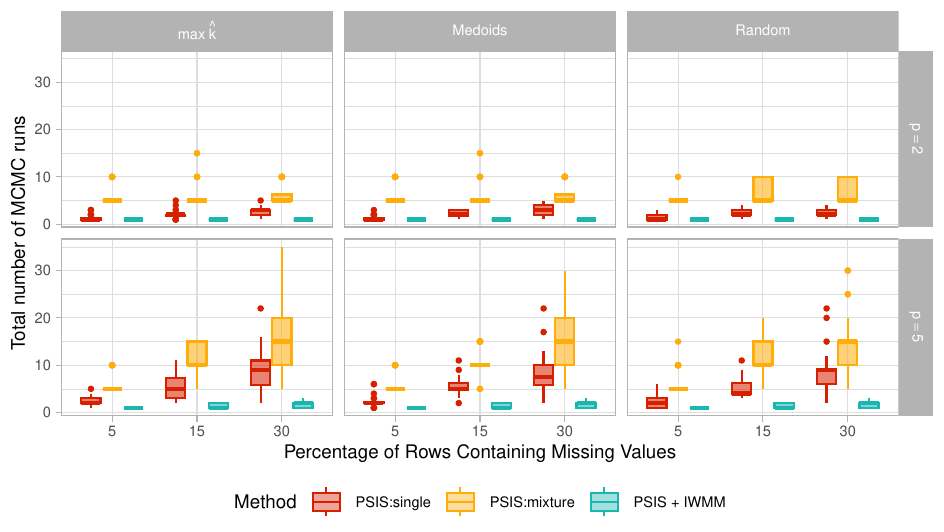}
    \caption{Number of MCMC runs necessary to approximate the posteriors of $m = 100$ imputed datasets with size $N=10$ and number of parameters $p$ (in the rows), using standard priors.  Different selection methods are employed (in the columns).}
    \label{fig:no_MCMC_n10}
\end{figure*}

\begin{figure*}[ht]
    \centering
    \includegraphics[width=\linewidth]{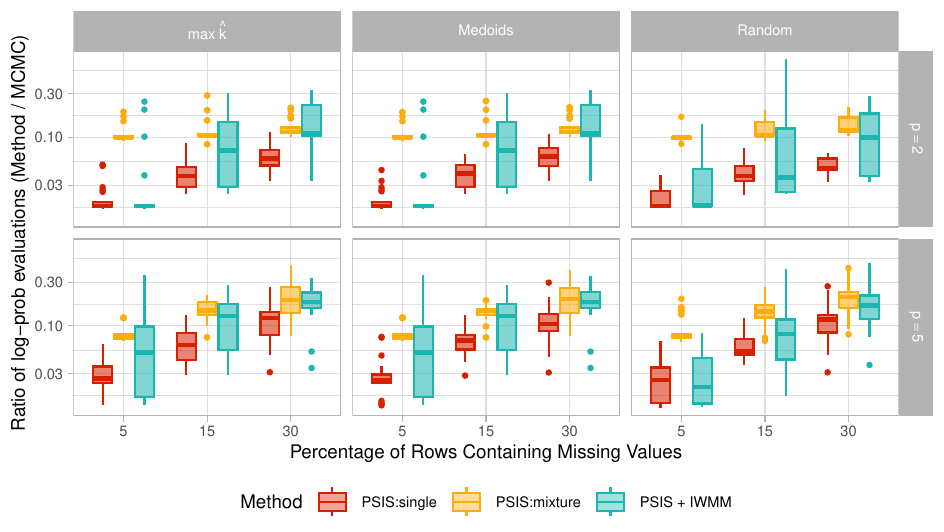}
    \caption{Ratio of number of evaluations of the log-probability compared to running MCMC separately necessary to approximate the posteriors of $m = 100$ imputed datasets with size $N=10$ and number of parameters $p$ (in the rows), using standard priors. Different selection methods are employed (in the columns).}
    \label{fig:ratio_log_lik_standard_n10}
\end{figure*}

\begin{figure*}[hb]
    \centering
    \includegraphics[width=\linewidth]{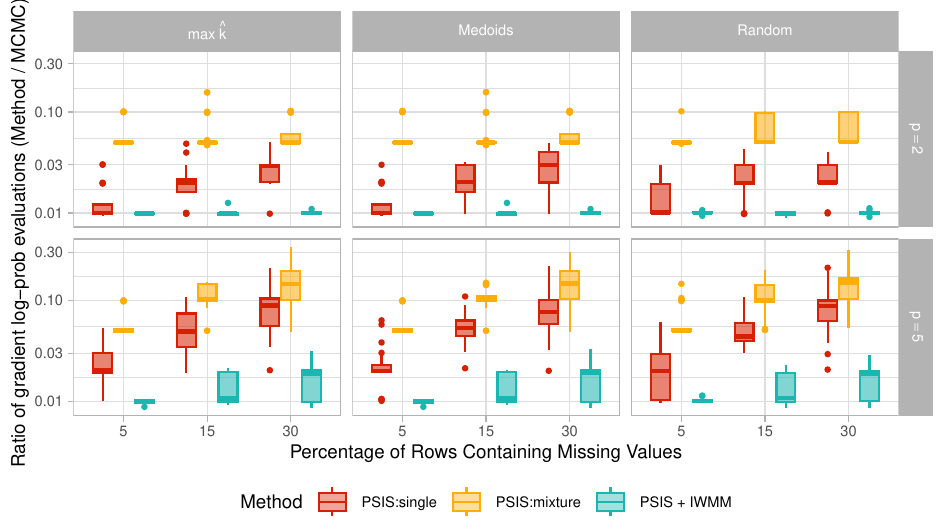}
    \caption{Ratio of number of evaluations of the gradient log-probability compared to running MCMC separately necessary to approximate the posteriors of $m = 100$ imputed datasets with size $N=10$ and number of parameters $p$ (in the rows), using standard priors. Different selection methods are employed (in the columns).}
    \label{fig:ratio_log_gradient_standard_n10}
\end{figure*}

\begin{figure*}[ht]
    \centering
    \includegraphics[width=\linewidth]{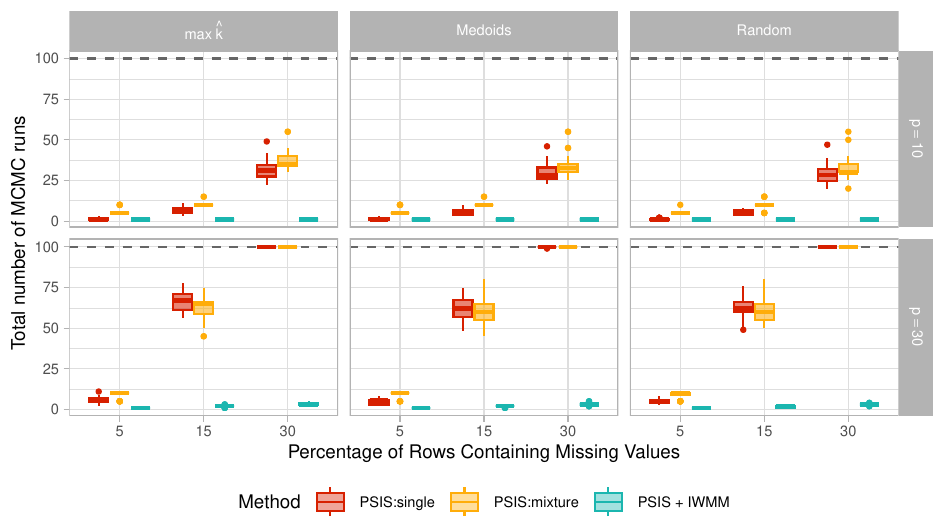}
    \caption{Number of MCMC runs necessary to approximate the posteriors of $m = 100$ imputed datasets with size $N=100$ and number of parameters $p$ (in the rows), using standard priors. Different selection methods are employed (in the columns).}
    \label{fig:no_MCMC_n100_standard}
\end{figure*}

\begin{figure*}[hb]
    \centering
    \includegraphics[width=\linewidth]{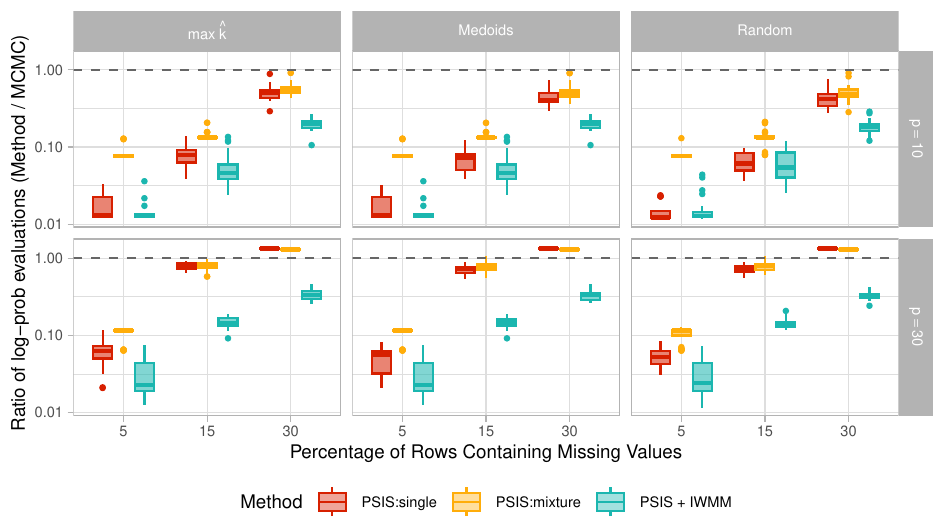}
    \caption{Ratio of number of evaluations of the log-probability compared to running MCMC separately necessary to approximate the posteriors of $ m = 100$ imputed datasets with size $N=100$ and number of parameters $p$ (in the rows), using standard priors. Different selection methods are employed (in the columns).}
    \label{fig:ratio_log_lik_standard_n100}
\end{figure*}

\begin{figure*}[ht]
    \centering
    \includegraphics[width=\linewidth]{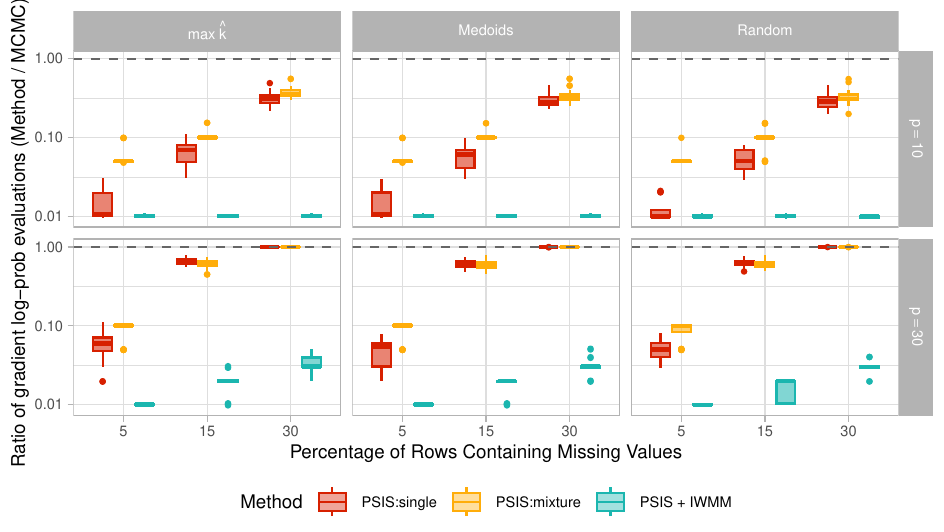}
    \caption{Ratio of number of evaluations of the gradient log-probability compared to running MCMC separately necessary to approximate the posteriors of $m = 100$ imputed datasets with size $N=100$ and number of parameters $p$ (in the rows), using standard priors. Different selection methods are employed (in the columns).}
    \label{fig:ratio_log_gradient_standard_n100}
\end{figure*}


\begin{figure*}[hb]
    \centering
    \includegraphics[width=\linewidth]{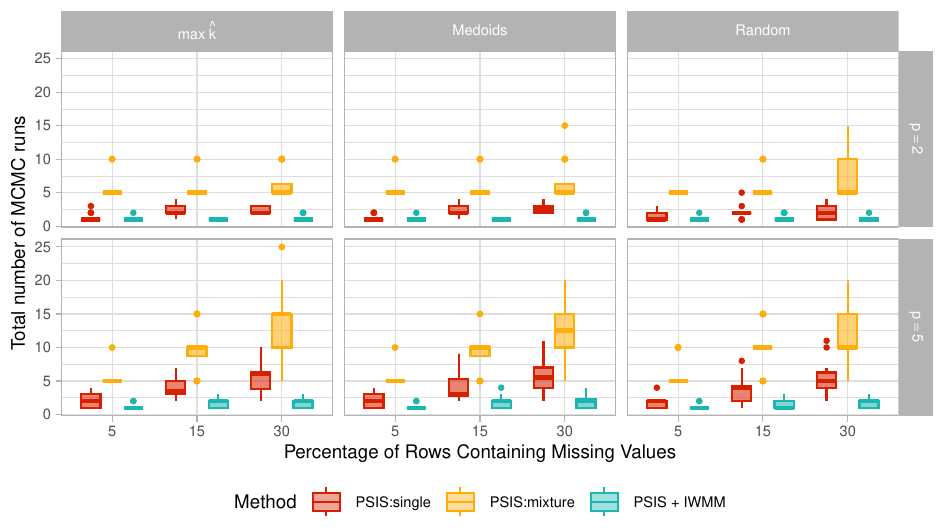}
    \caption{Number of MCMC runs necessary to approximate the posteriors of $m = 100$ imputed datasets with size $N=10$ and number of parameters $p$ (in the rows), using horseshoe priors. Different selection methods are employed (in the columns).}
    \label{fig:no_MCMC_n10_hs}
\end{figure*}

\begin{figure*}[ht]
    \centering
    \includegraphics[width=\linewidth]{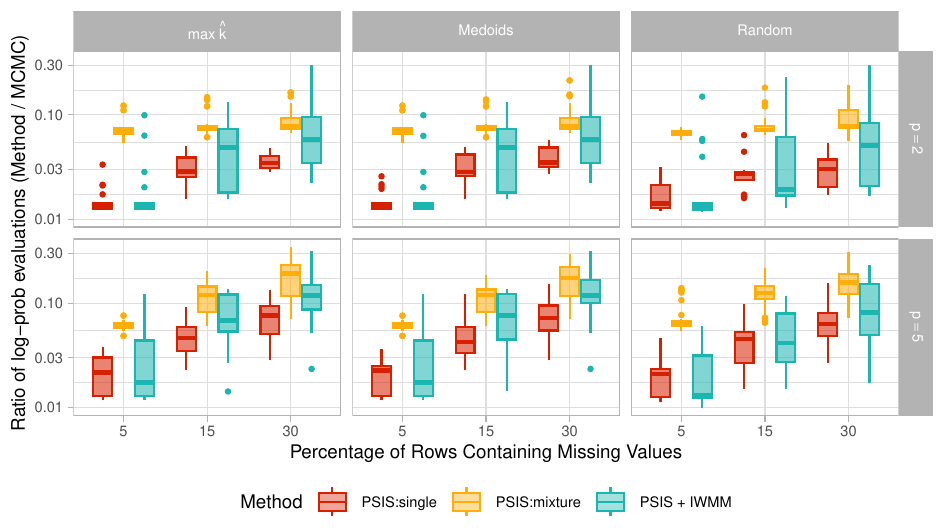}
    \caption{Ratio of number of evaluations of the log-probability compared to running MCMC separately necessary to approximate the posteriors of $m = 100$ imputed datasets with size $N=10$ and number of parameters $p$ (in the rows), using horseshoe priors. Different selection methods are employed (in the columns).}
    \label{fig:ratio_log_lik_hs_n10}
\end{figure*}

\begin{figure*}[hb]
    \centering
    \includegraphics[width=\linewidth]{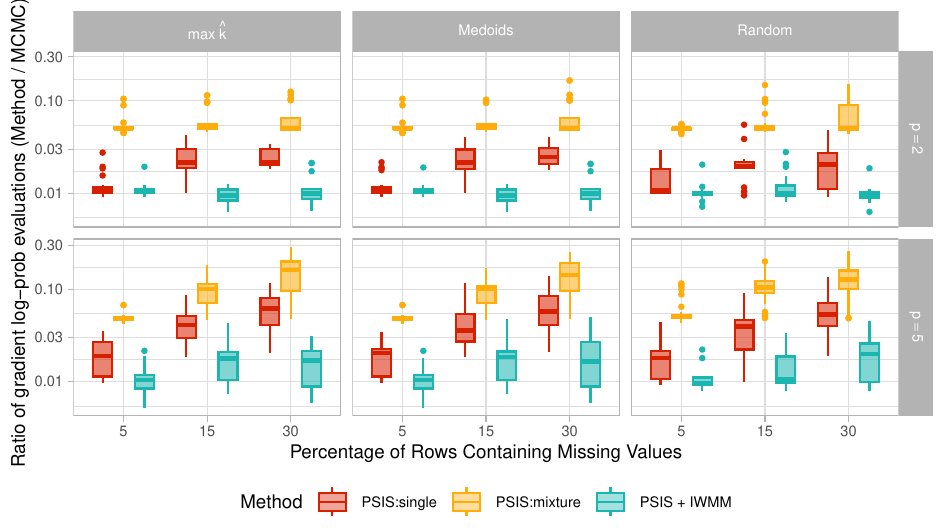}
    \caption{Ratio of number of evaluations of the gradient log-probability compared to running MCMC separately necessary to approximate the posteriors of $m = 100$ imputed datasets with size $N=10$ and number of parameters $p$ (in the rows), using horseshoe priors. Different selection methods are employed (in the columns).}
    \label{fig:ratio_log_gradient_hs_n10}
\end{figure*}

\begin{figure*}[ht]
    \centering
    \includegraphics[width=\linewidth]{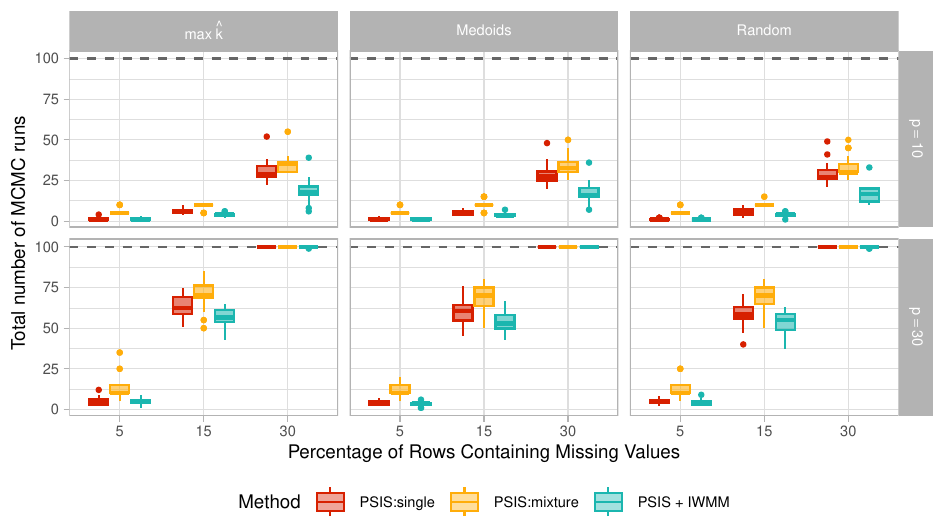}
    \caption{Number of MCMC runs necessary to approximate the posteriors of $m = 100$ imputed datasets with size $N=100$ and number of parameters $p$ (in the rows), using horseshoe priors. Different selection methods are employed (in the columns).}
    \label{fig:no_MCMC_n100_hs}
\end{figure*}

\begin{figure*}[hb]
    \centering
    \includegraphics[width=\linewidth]{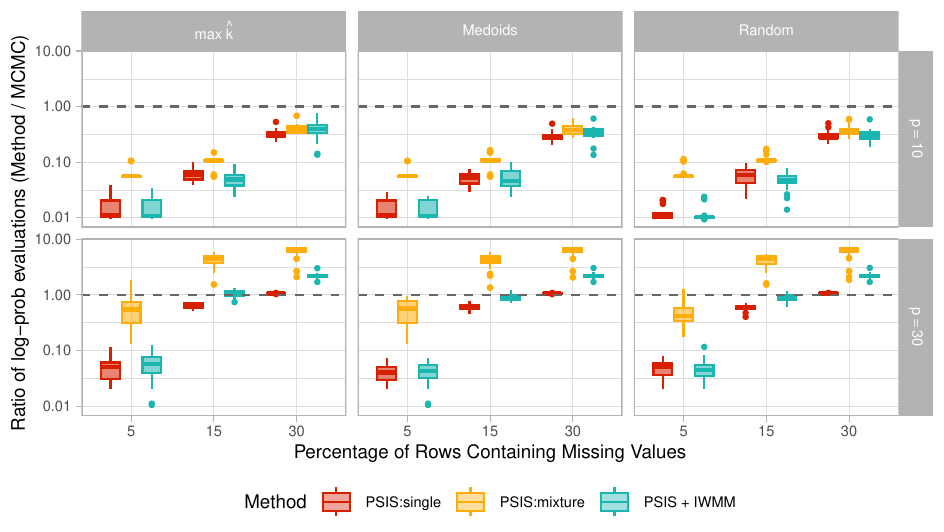}
    \caption{Ratio of number of evaluations of the log-probability compared to running MCMC separately necessary to approximate the posteriors of $m = 100$ imputed datasets with size $N=100$ and number of parameters $p$ (in the rows), using horseshoe priors. Different selection methods are employed (in the columns).}
    \label{fig:ratio_log_lik_hs_n100}
\end{figure*}

\begin{figure*}[ht]
    \centering
    \includegraphics[width=\linewidth]{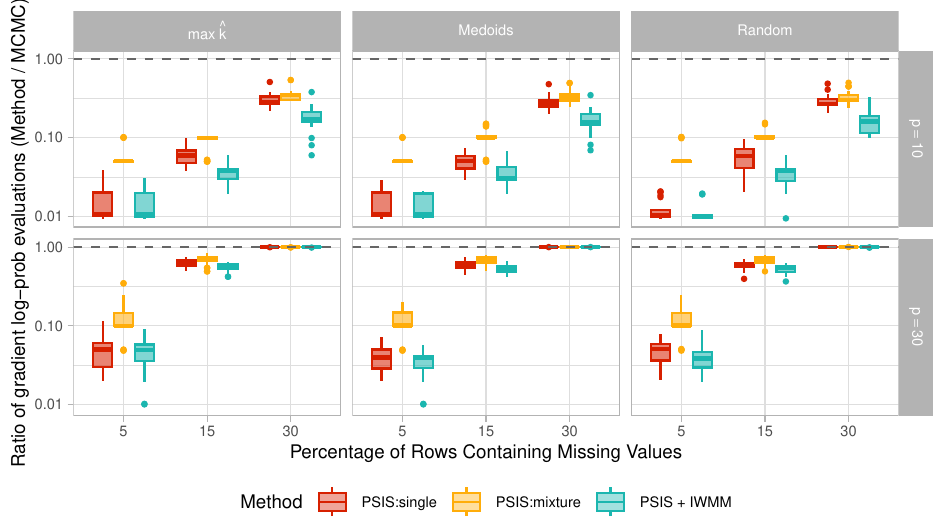}
    \caption{Ratio of number of evaluations of the gradient log-probability compared to running MCMC separately necessary to approximate the posteriors of $m = 100$ imputed datasets with size $N=100$ and number of parameters $p$ (in the rows), using horseshoe priors. Different selection methods are employed (in the columns).}
    \label{fig:ratio_log_gradient_hs_n100}
\end{figure*}

\FloatBarrier
\onecolumn
\section{Additional Results of the Simulation Study for the Surrogate Modeling Problem \label{sec:ap:surrogate_sim_results}}

\begin{figure*}[ht]
    \centering
    \includegraphics[width=\linewidth]{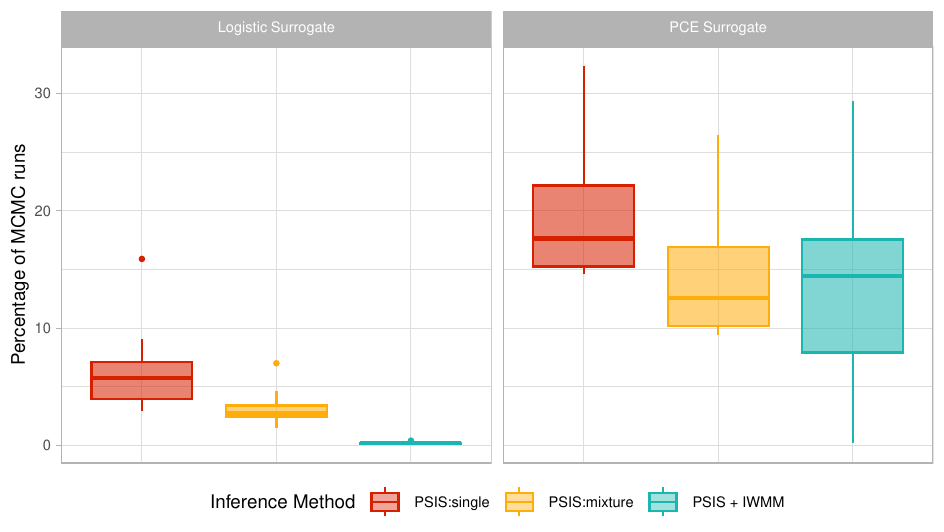}
    \caption{Percentage of MCMC runs required to approximate the posteriors of $m=1000$ surrogate parameter draws, for the logistic and PCE surrogate (in columns).}
    \label{fig:surrogate_perc_MCMC_1000}
\end{figure*}

\begin{figure*}[h]
    \centering
    \includegraphics[width=\linewidth]{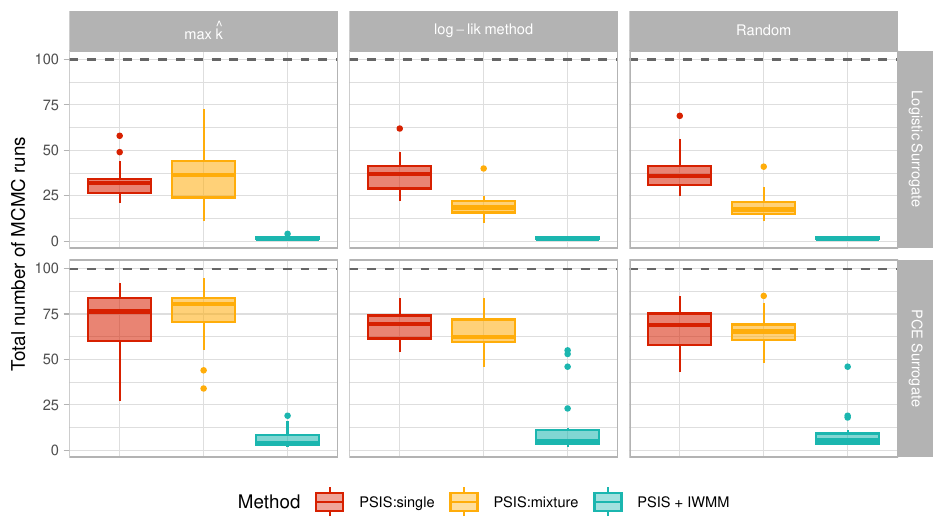}
    \caption{Number of MCMC runs required to approximate the posteriors of $m=100$ surrogate parameter draws, for the logistic and PCE surrogate (in rows) and different selection methods (in columns).}
    \label{fig:surr_no_MCMC_select}
\end{figure*}

\begin{figure*}
    \centering
    \includegraphics[width=\linewidth]{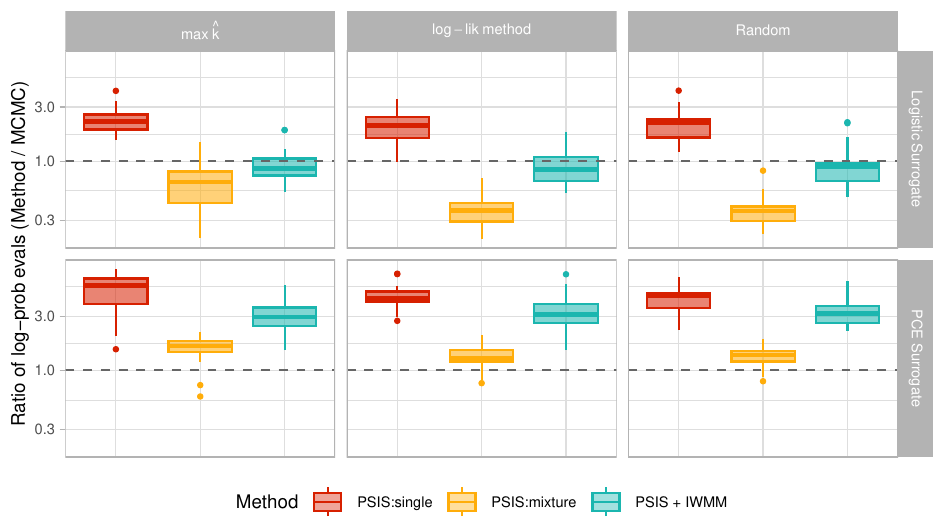}
    \caption{Ratio of number of evaluations of the log-probability (on a log-scale) compared to running MCMC separately necessary to approximate the posteriors of $m = 100$ surrogate parameter draws, for the logistic and PCE surrogate (in rows) and different selection methods (in columns).}
    \label{fig:surr_log_ratio_select}
\end{figure*}

\begin{figure*}
    \centering
    \includegraphics[width=\linewidth]{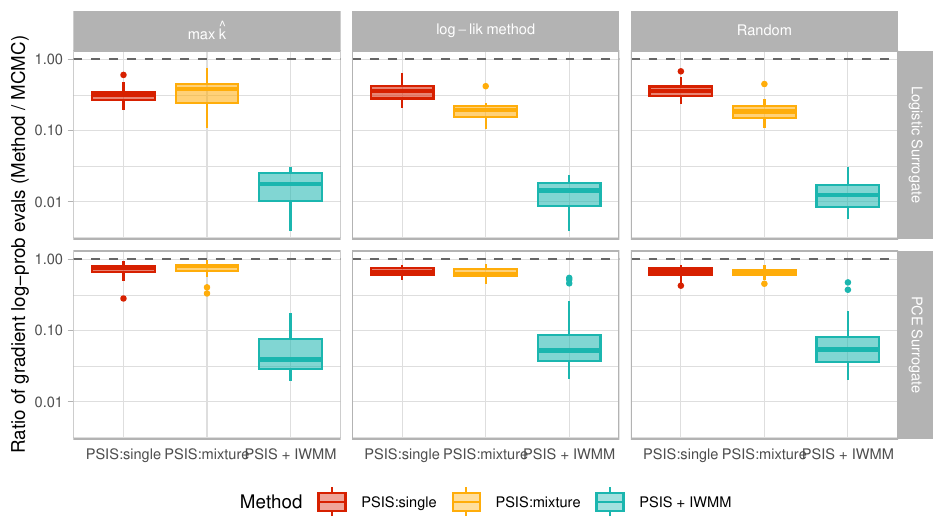}
    \caption{Ratio of number of evaluations of the gradient log-probability (on a log-scale) compared to running MCMC separately necessary to approximate the posteriors of $m = 100$ surrogate parameter draws, for the logistic and PCE surrogate (in rows) and different selection methods (in columns).}
    \label{fig:surr_gradient_ratio_select}
\end{figure*}

\end{appendices}
\FloatBarrier
\twocolumn
\bibliography{references}

\end{document}